\documentclass[prb,preprint,superscriptaddress]{revtex4-1}

\usepackage{tabularx,booktabs}
\usepackage{multirow}
\usepackage{amsmath}
\usepackage{amssymb}
\usepackage{graphicx}
\graphicspath{{figs/}}
\usepackage{hyperref}
\hypersetup{
 bookmarks=true,		
 unicode=false,			
 pdftoolbar=true,		
 pdffitwindow=false,		
 pdfstartview={FitH},		
 pdfcreator={pdflatex},		
 pdfproducer={vim},		
 pdfnewwindow=true,		
 colorlinks=true,		
 linktoc=page,			
 linkcolor=blue,		
 citecolor=blue,		
 filecolor=blue,		
 urlcolor=blue			
}

\begin{document}
\title{First principles theory of the nitrogen interstitial in hBN: a plausible model for the blue emitter}
\date{\today}
\author{Ádám Ganyecz}
\affiliation{Strongly Correlated Systems Lend\"{u}let Research Group, Wigner Research Centre for Physics, H-1525, Budapest, Hungary}
\affiliation{MTA–ELTE Lend\"{u}let "Momentum" NewQubit Research Group, Pázmány Péter, Sétány 1/A, 1117 Budapest, Hungary}
\author{Rohit Babar}
\affiliation{Strongly Correlated Systems Lend\"{u}let Research Group, Wigner Research Centre for Physics, H-1525, Budapest, Hungary}
\affiliation{MTA–ELTE Lend\"{u}let "Momentum" NewQubit Research Group, Pázmány Péter, Sétány 1/A, 1117 Budapest, Hungary}
\author{Zsolt Benedek}
\affiliation{Strongly Correlated Systems Lend\"{u}let Research Group, Wigner Research Centre for Physics, H-1525, Budapest, Hungary}
\affiliation{MTA–ELTE Lend\"{u}let "Momentum" NewQubit Research Group, Pázmány Péter, Sétány 1/A, 1117 Budapest, Hungary}
\author{Igor Aharonovich}
\affiliation{School of Mathematical and Physical Sciences, Faculty of Science, University of Technology Sydney, Ultimo, New South Wales 2007, Australia}
\affiliation{ARC Centre of Excellence for Transformative Meta-Optical Systems (TMOS), Faculty of Science, University of Technology Sydney, Ultimo, New South Wales 2007, Australia}
\author{Gergely Barcza}
\email{barcza.gergely@wigner.hu}
\affiliation{Strongly Correlated Systems Lend\"{u}let Research Group, Wigner Research Centre for Physics, H-1525, Budapest, Hungary}
\affiliation{MTA–ELTE Lend\"{u}let "Momentum" NewQubit Research Group, Pázmány Péter, Sétány 1/A, 1117 Budapest, Hungary}
\author{Viktor Ivády}
\email{ivady.viktor@ttk.elte.hu}
\affiliation{MTA–ELTE Lend\"{u}let "Momentum" NewQubit Research Group, Pázmány Péter, Sétány 1/A, 1117 Budapest, Hungary}
\affiliation{Department of Physics of Complex Systems, Eötvös Loránd University, Egyetem tér 1-3, H-1053 Budapest, Hungary}
\affiliation{Department of Physics, Chemistry and Biology, Link\"oping University, SE-581 83 Link\"oping, Sweden}

\begin{abstract}
Color centers in hexagonal boron nitride (hBN) have attracted considerable attention due to their remarkable optical properties enabling robust room temperature photonics and quantum optics applications in the visible spectral range. On the other hand, identification of the microscopic origin of color centers in hBN has turned out to be a great challenge that hinders in-depth theoretical characterization, on-demand fabrication, and development of integrated photonic devices. This is also true for the blue emitter, which is an irradiation damage in hBN emitting at 436~nm wavelengths with desirable properties. Here, we propose the negatively charged  nitrogen split interstitial defect in hBN as a plausible microscopic model for the blue emitter. To this end, we carry out a comprehensive first principles theoretical study of the nitrogen interstitial. We carefully analyze the accuracy of first principles methods and show that the commonly used HSE hybrid exchange-correlation functional fails to describe the electronic structure of this defect. Using the generalized Koopman's theorem, we fine tune the functional and obtain a zero-phonon photoluminescence (ZPL) energy in the blue spectral range. We show that the defect exhibits high emission rate in the ZPL line and features a characteristic phonon side band that resembles the blue emitter's spectrum. Furthermore, we study the electric field dependence of the ZPL and numerically show that the defect exhibits a quadratic Stark shift for perpendicular to plane electric fields, making the emitter insensitive to electric field fluctuations in first order. Our work emphasize the need for assessing the accuracy of common first principles methods in hBN and exemplifies a workaround methodology.  Furthermore, our work is a step towards understanding the structure of the blue emitter and utilizing it in photonics applications.
\end{abstract}

\maketitle

\section{Introduction}

The development of quantum communication\cite{Gisin_2007,Orieux_2016,Cozzolino_2019}   and quantum internet\cite{kimble_quantum_2008,wehner_quantum_2018} demand the preparation and processing of tailored photonic states to carry pieces of quantum information to long distances.\cite{pelucchi_potential_2022} Complex photonic technologies require on-demand emission of single photons from controllable single photon emitters (SPEs). For real-life applications SPEs need to produce high purity single photons with a high emission rate, be robust against environmental disturbances, and should offer solutions for chip-scale integration in photonic devices.

Numerous material platforms, ranging from silicon through diamond to polymers, have been proposed and tested for integrated quantum photonic applications.\cite{pelucchi_potential_2022} Each platform offers different advantages, however, so far no material has been found that can satisfy all the needs for large scale photonic applications. Therefore, currently the most viable approach is the optimization of different processing elements in various hosts materials offering cross-platform integration capabilities. \cite{pelucchi_potential_2022}

On the forefront of single photon emitter developments, color centers in hexagonal boron nitride (hBN) have recently gained considerable attention.\cite{caldwell_photonics_2019}  Numerous color centers have been reported that feature sharp ZPL line at room temperature, on demand single photon emission rate in the MHz range, high spectral stability, and high quantum efficiency.\cite{aharonovich_quantum_2022,kubanek_coherent_2022} These color centers can be engineered to realize SPEs emitting at various wavelengths in a wide spectral range.\cite{aharonovich_quantum_2022,kubanek_coherent_2022} Due to its layered structure, hBN also offers versatile integration capabilities for photonic applications.\cite{caldwell_photonics_2019,al-juboori_quantum_2023}

Very recently, a color center emitting at $\sim$436~nm in hBN, named as the blue emitter, has shown remarkable properties.\cite{shevitski_blue-light-emitting_2019,caldwell_photonics_2019,fournier_position-controlled_2021,liang_blue_2023} In addition to the sharp room temperature emission line and high emission rate, the blue emitter is known for its outstanding spectral stability and robustness.\cite{fournier_position-controlled_2021,horder_coherence_2022,zhigulin_stark_2022,higulin_photophysics_2023,liang_blue_2023} The latter features are attributed to a quadratic electric field dependence of the Stark shift of the blue emitter, which make the defect insensitive to weak electric field fluctuations in first order.\cite{zhigulin_stark_2022}

The blue emitter appears as a radiation damage in hBN and it is often observed in irradiated samples.\cite{shevitski_blue-light-emitting_2019,fournier_position-controlled_2021,gale_site-specific_2022,liang_blue_2023} In addition, the appearance of the blue emitter seems to show a close relation to another emitter in the ultraviolet (UV) spectral range\cite{gale_site-specific_2022}. The microscopic structure of the UV emitter is currently debated in the literature.\cite{mackoit-sinkeviciene_carbon_2019,hamdi_stonewales_2020,li_ultraviolet_2022} This is also true for the blue emitter, as the identification of the blue emitter's microscopic structure remains elusive hindering theoretical studies and slowing down the development of related photonic applications.\cite{liang_blue_2023}

From the analysis of the blue emitter's photophysics, several important properties of the underlying atomic and electronic structure have been deduced.\cite{zhigulin_stark_2022,zhigulin_photophysics_2023} From the quadratic Stark effect one could infer a high, D$_{3h}$ symmetry of the ground and excited state configurations.\cite{zhigulin_stark_2022} In addition, the  stability of the center under continuous optical pumping suggests a closed shell ground state with fully occupied and completely empty defect state(s) in the lower half and the upper half of the band gap, respectively.\cite{zhigulin_photophysics_2023} Taking into considerations the conditions of the fabrication, vacancies, interstitials, antisites, and complexes derived from the combination of these defects are the most plausible candidates of the microscopic structures that can be created by irradiation in hBN.\cite{liang_blue_2023}  Related structures, such as the nitrogen interstitial and the carbon split interstitial dimer, have already been suggested in Ref.~[\onlinecite{zhigulin_stark_2022}] as potential candidates that fulfill the above mentioned criteria.  Furthermore, these defects were also discussed in several recent theoretical works.\cite{Wang2014def,weston_native_2018,Strand2019,khorasani_identification_2021,bhang_first-principles_2021} On the other hand, no comprehensive theoretical studies have been carried out for these candidates.

Here, we theoretically study the negatively charged nitrogen interstitial defect and propose it as a microscopic model for blue emitter in hBN. So far, no consensus has been achieved in the literature concerning the most favourable configuration of this defect. We show that the split interstitial configuration is the lowest energy form of the nitrogen interstitial. We carefully analyze the accuracy of popular density functional theory (DFT) methods by testing the generalized Koopman's theorem\cite{lany_polaronic_2009,lany_generalized_2010, ivady_role_2013} for the defect states and carrying out fractional occupation number weighted electron density (FOD)\cite{grimme2015practicable,bauer2017fractional} analysis on the system. We show that the excited state of the defect, exhibiting stretched nitrogen-boron bonds, cannot be described by the hybrid functional most often used for hBN. The fraction of the exact exchange contribution needs to be reduced to obtain a higher accuracy ZPL value. Furthermore, we calculate the photoluminescence (PL) spectra of the negatively charged nitrogen split interstitial and show that it resembles the experimental spectra of the blue emitter. We also study the electric field dependence of the ZPL energy and numerically demonstrate a quadratic Stark shift. We conclude that the nitrogen split interstitial defect is a promising model for the blue emitter which may facilitate advances in the fabrication and utilization of the emitter.

\section{Results - First principles theory of the nitrogen interstitial in hBN}

In this section, we carry out a focused theoretical study on the nitrogen split interstitial defect in hBN. Our goal is to achieve the highest possible accuracy for the N$_i$ defect by utilizing a diverse set of state-of-the-art numerical methods. 
Relevant details of the methodology are incorporated into the discussions of the results in this section. The remaining technical details of the calculations are provided in the Methods section.

\subsection{The stable atomic configuration}

First, we study the stable and metastable atomic configurations of an inter-layer nitrogen interstitial in hBN. We consider the negative charge state of the system, which is expected to be the most stable for Fermi energy values close to the middle of the band gap.\cite{weston_native_2018, khorasani_identification_2021}  Various related structures have been studied in the literature before;\cite{Wang2014def,weston_native_2018,Strand2019,khorasani_identification_2021} however, no consensus has been achieved regarding the ground state configuration.
Although multiple studies identify the [0001] split interstitial as the most favored configuration,\cite{Wang2014def,weston_native_2018,Strand2019} a recent study found a tilted split interstitial configuration for the ground state.\cite{khorasani_identification_2021} Furthermore, this tilted split interstitial configuration was proposed as a candidate for the 3.1 eV emitter.\cite{khorasani_identification_2021}

\begin{figure*}[!th]
\begin{center}
\includegraphics[width=0.9\textwidth]{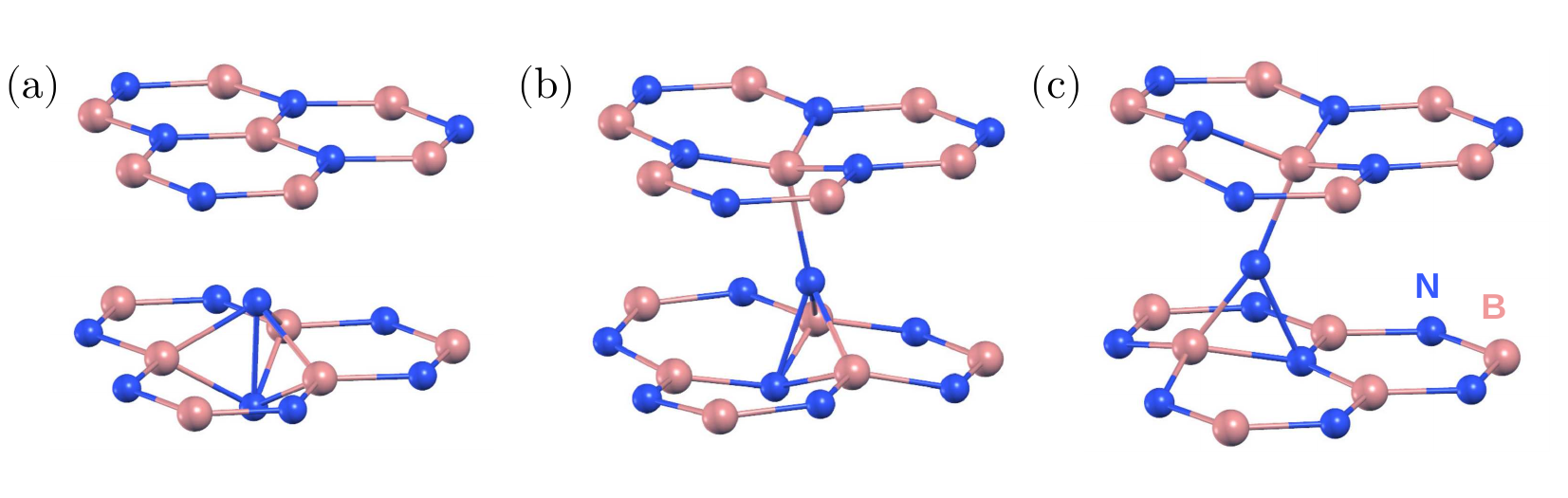}
	\caption{Atomic structure of negatively charged nitrogen interstitial defect in hBN. (a) The split interstitial with interlayer bonding is the energetically favored configuration. We note that there is no covalent bond between the two nitrogen atoms of the split interstitial defect. (b) and (c) denote the metastable configurations with intralayer bonding, and formation energies 0.51 eV and 0.50 eV (0.50 eV and 0.59 eV) calculated using HSE(0.3) (PBE) higher than the split interstitial, respectively. Nitrogen (boron) atoms are shown as blue (pink) spheres.}
	\label{fig:fig1_struc}  
	\end{center}
\end{figure*}

In order to unambiguously identify the ground state configuration of the nitrogen interstitial in hBN, we investigate a series of potentially relevant defect configurations in a multi-layered bulk systems, see Fig.~\ref{fig:fig1_struc}. We first place the interstitial atom into a small orthorhombic 120-atom supercell of two layer hBN with AA$^{\prime}$ stacking similar to the model used in Ref.[\onlinecite{khorasani_identification_2021}]. Since relaxation of the defect and over-mixing of defect and host states can lead to different ground state structures\cite{Gouveia2019}, we consider both semi-local PBE functional\cite{PBE} and hybrid HSE functional with Fock exchange contribution of $\alpha = 0.30$ and screening parameter $\mu$ set to 0.4 \AA$^{-1}$ (also used in Ref.~[\onlinecite{khorasani_identification_2021}]) to identify the ground state. We find the [0001] split interstitial, Fig.~\ref{fig:fig1_struc}(a), to be favorable by 0.59 eV (0.68 eV) than the lowest energy tilted split-interstitial configuration calculated using HSE(0.3) (PBE) functional. Therefore, we could not confirm the results of Ref.~[\onlinecite{khorasani_identification_2021}], but our results are in-line with earlier reports of Refs[\onlinecite{Wang2014def,weston_native_2018,Strand2019}] 

Before discussing our results for larger, convergent supercells, we point out shortcomings of the two-layer model. To study the stability of the model, we carry out molecular dynamics (MD) simulation at 500 K using PBE functional. As a result of the atomic thermal motion, we observe sliding of adjacent BN layers from stacking AA$^{\prime}$ to AB1$^{\prime}$, where the B atoms are aligned on top of each other, while N atoms slide to the hollow site.\cite{Gilbert2019} The fully relaxed [0001] split interstitial in the AB1$^{\prime}$ stacked hBN is found to be 0.68 eV (0.95 eV) more favorable than in the AA$^{\prime}$ stacked hBN calculated by using HSE(0.3) (PBE) functional. This is due to reduction of the the repulsive interaction between the N atoms in the AB1$^{\prime}$ stacking. From the analysis of the atomic configurations proposed in Ref.~[\onlinecite{khorasani_identification_2021}] and from our experiences with two-layer models, we conclude that the recently reported tilted ground state  configuration\cite{khorasani_identification_2021} is possibly a consequence of unphysical layer slippage. We note here that the sliding of the BN layers results in a metastable configuration in a four-layer model. To avoid slippage of the layers in the simulations, we include at least 4 layers in our bulk models hereinafter. 

According to our findings, the hexagonal 512-atom supercell comprising of four BN layers is sufficient to accurately describe the ground state configurations of the nitrogen interstitial in hBN. Using HSE(0.3) functional and the 512-atom model, the [0001] split interstitial configuration is found to be the most favorable one, see Fig.~\ref{fig:fig1_struc}(a), followed by a tilted interstitial configuration with tilt angle of 17\textdegree{} from the [0001] axis along the hollow region and 0.51 eV higher in energy (16\textdegree{} and 0.50 eV with PBE), see Fig.~\ref{fig:fig1_struc}(b). Another metastable interstitial configuration is obtained with 0.50 eV higher energy and pronounced tilt of 26\textdegree{} in the BN bond direction (25\textdegree{} and 0.59 eV with PBE). Unlike the previous configurations, the 26\textdegree{} tilted interstitial consists of three-coordinated N  atom, see Fig.~\ref{fig:fig1_struc}(c). Additional calculations indicate that the energy difference between different configurations is more sensitive to the exchange fraction in comparison to separation between hBN layers and choice of van der Waals screening.

Using HSE(0.3) functional, the distance between the split interstitial N atoms and  the neighboring B atoms are $d_1$ (interstitial nitrogen-nitrogen pair) 1.56~\AA, $d_2$ (interstitial nitrogen with closest boron in same layer) 1.57~\AA, and $d_3$ (interstitial nitrogen with boron from top/bottom layer) 2.57~\AA. The PBE distance are $d_1 = 1.60$~\AA, $d_2 = 1.58$~\AA, $d_3 = 2.56$~\AA. 
The negatively charged split interstitial defect exhibits high, D$_{3h}$ symmetry in the ground state.

\subsection{Electronic structure}

\begin{figure*}[!th]
\includegraphics[width=0.99\textwidth]{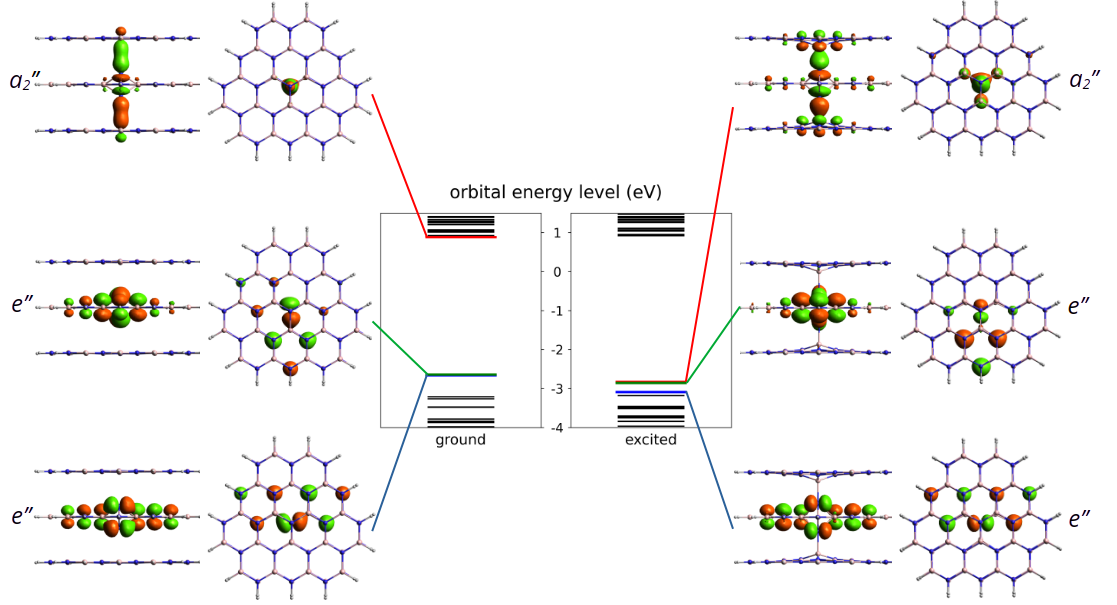}
\caption{Kohn-Sham energy levels of the ground and excited states. LUMO, HOMO and HOMO-1 frontier orbital energies are highlighted in red, green and blue, respectively. The shape of the corresponding frontier orbitals are also plotted in both side and top view. Note that for better visibility only the central three layers of the molecules are shown.
\label{fig:energylevels}
}
\end{figure*}

After identifying the most favourable configuration of the negatively charged nitrogen split interstitial defect in hBN, we continue our study with the analysis of the electronic structure of the defect. The two out-of-plane nitrogen atoms give rise to several defect states, some of which fall into the valence band. The ones that appear in the band gap are the fully occupied $e^{\prime\prime}$ state and the empty $a_2^{\prime\prime}$ state, see Fig.~\ref{fig:energylevels}. The ground state is a singlet, and the defect possesses no spin.

Optical transition between the occupied and the unoccupied defect states is allowed by perpendicular to $c$ polarized photons. In the excited state, the nitrogen atoms form long bonds with the closest boron atoms on the symmetry axis due to the occupation of the bonding $a_2^{\prime\prime}$ state. Consequently, the N-N distance $d_1$ extends by about 26\% (0.42~\AA\ ), while the B-N distance $d_3$ on the symmetry axis contract by 33\% (0.85~\AA\ ). We note that the excited state of the N$_i$(-) defect is slightly distorted due to the Jahn-Teller effect that reduces the symmetry to C$_{2v}$. The three B-N distances in the middle plane ($d_2$) change from 1.58~\AA, 1.58~\AA, and 1.58~\AA\ to 1.67~\AA, 1.67~\AA, and 1.58~\AA.

\begin{figure*}[!t]
\centering
\begin{tabular}[t]{cccccc}
{\bf a} & & {\bf b} & & {\bf c}\\
&\includegraphics[width=0.36\textwidth]{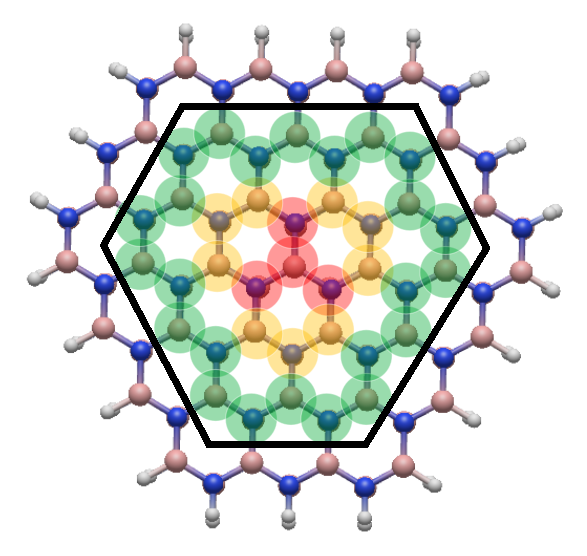}&
 &\includegraphics[width=0.28\textwidth]{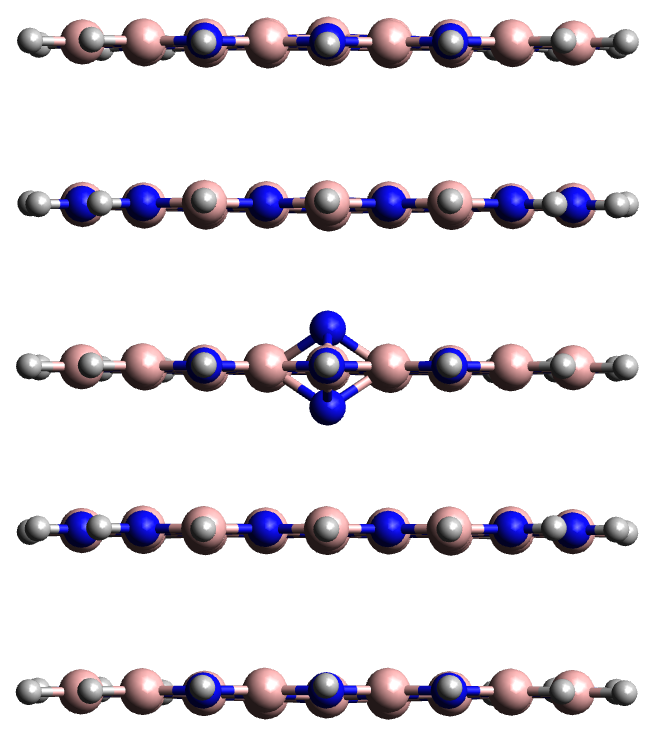}&
 & \includegraphics[width=0.28\textwidth]{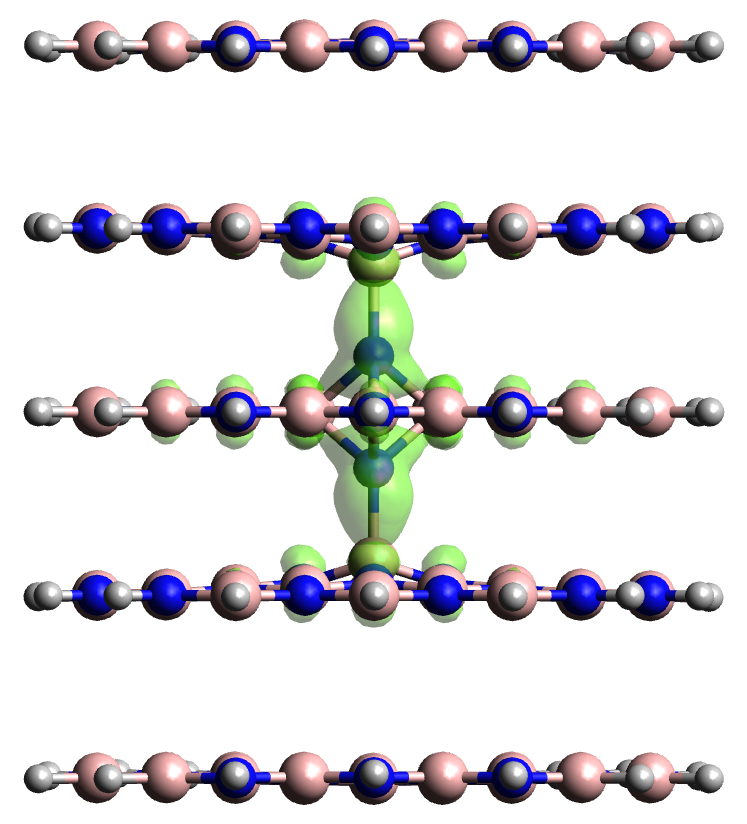}\\
\end{tabular}
\caption{\label{fig:structure_fod}
\textbf{a}, shows the structure of the larger layer from top view. Note that the second-row atoms included in the smaller layer are enclosed by the black hexagon while the grey, pink and blue balls denote the position of the hydrogen, boron and nitrogen atoms, respectively. The translucent red, yellow, and green highlighting represents the minimal and the additional subsets of atoms included in molecular geometry relaxation, respectively. For more details see main text.  
FOD analysis for the ground state and the excited state geometry presented in \textbf{b} and \textbf{c} subplot, respectively.  Translucent green denotes the FOD surface at 0.005 e/Bohr$^3$ and T=5000 K.
}
\end{figure*}

Accurate first principles calculation of the optical properties of color centers in hBN has turned out to be a challenge. In order to draw reliable conclusions for the N$_i$(-) defect, we test different methods, such as PBE and HSE($\alpha$) functionals in periodic supercell models and PBE-TDDFT and CASSCF-NEVPT2 methods on cluster models of the N$_i$(-) defect in hBN. In periodic 768-atom models, we use the constrained occupation method and Hellmann-Feynman forces to relax the excited state structure corresponds to $e^{\prime\prime} \rightarrow a_{2}^{\prime\prime}$ transition. Excited states geometry is first relaxed with PBE functional and then continued with HSE($\alpha$) functional. We note that relaxation of the excited state with HSE($\alpha$)  ($\alpha \approx 0.25$ - $0.35$) is cumbersome due to the change of ordering of the defect states and related structural instabilities. In our 261-atom cluster model of 5 adjacent hBN flakes, see Fig.~\ref{fig:structure_fod}, we initially optimize the ground state geometry at PBE level of theory, using def2-SVP basis set. During the optimization, owing to its computational demands, only the middle 13 atoms are relaxed in each layer plus the additional interstitial nitrogen. On the obtained equilibrium geometry, we calculate the vertical excitation spectrum with PBE-TDDFT and CASSCF-NEVPT2 methods, requesting 3 roots (ground state and 2 degenerate excited states in all cases).
Then, we relax the structure of the excited state by PBE-TDDFT, following the energy gradient of the corresponding root. The electronic energy of the excited state is calculated from a second vertical excitation spectrum calculation at the relaxed geometry. 
Using a larger basis set would have been too expensive, however to obtain more accurate ZPL, def2-TZVPD basis is applied on the middle 4 atoms of each layer plus the interstitial nitrogen while def2-SVP basis set on the remaining atoms.
More details on the excited state calculations and the models can be found in the Methods section.

The adiabatic energy differences ($E_{\text{AD}} = E_{\text{ES}} - E_{\text{GS}}$) calculated  for the various models are provided in Table~\ref{tab:ZPL}. The true ZPL energy includes contribution from zero point energies of the phonon modes in the ground and excited states ( $E_{\text{ZPL}} = E_{\text{ES}} - E_{\text{GS}} + \Delta E_{\text{ZPE}}$, where $\Delta E_{\text{ZPE}} = E_{\text{ZPE}}^{\text{ES}}-E_{\text{ZPE}}^{\text{GS}}$), see later. As can be seen, the adiabatic energy differences ranges in a wide energy interval starting at 2.38~eV up to 3.20~eV. The wavefunction-based NEVPT2 method predicts the highest energy difference of 3.20~eV.

\begin{table}[!h]
\begin{center}
\caption{\label{tab:ZPL} Adiabatic energy differences and ZPL values as obtained with different computational methods. The values in parentheses are given in nm, while all the other values are in eV. The zero point energy contribution (ZPE) contribution to the adiabatic energy difference is -0.10~eV. Bold font indicates our most accurate theoretical model, adiabatic energy difference, and  ZPL.} 
 \begin{tabular}{c|c|c|c  }
 \hline
\multirow{3}{*}{Model} & \multirow{3}{*}{Method} & Adiabatic energy  & \multirow{2}{*}{ZPL}  \\ 
&  & difference &  \\
&  & $E_{\text{AD}}$ & $E_{\text{ZPL}} = E_{\text{AD}} + E_{\text{ZPE}}   $\\ \hline
261-atom cluster & CASSCF(4,3)-NEVPT2 & 3.20 & 3.10 \\
261-atom cluster & PBE-TDDFT & 2.38 &  2.28\\
768-atom supercell & PBE & 2.40 & 2.30 \\ \hline
768-atom supercell & HSE(0.32) & 3.33 & 3.23 \\
768-atom supercell & HSE(0.285) & 3.27 & 3.18 \\
\textbf{768-atom supercell} & \textbf{HSE(0.208)} & \textbf{3.06} & \textbf{2.96} \\
768-atom supercell & HSE(0.132) & 2.84 & 2.74 \\ \hline
\multicolumn{3}{c|}{Experiment} &  2.844 (436)\cite{zhigulin_stark_2022} \\ \hline
 \end{tabular}
 \end{center}
\end{table}

As demonstrated for other systems\cite{ivady_ab_2020,babar_quantum_2021,benedek_symmetric_2023}, the CASSCF-NEVPT2 method can provide accurate results when excited states of multi-reference character are considered. However, for the N$_i$(-) defect, both the ground state and the lowest energy singlet excited state possess a clear single-reference character. The leading electron configurations visualized in Fig.~\ref{fig:energylevels} have 100.0\% and 98.9\% contribution in the ground and excited state wavefunctions, respectively, if 4 electrons and the 3 frontier orbitals form the active space. We note that similar wavefunction characters are obtained for systematically extended active spaces. Therefore, the accuracy of the NEVPT2 method in this case is comparable to the MP2 method\cite{Angeli-2001a}, which is expected to be outperformed by a carefully chosen single-determinant DFT method.

 As an alternative investigation of the reliability of single-reference methods, we adopt the fractional occupation number weighted electron density (FOD) analysis method~\cite{grimme2015practicable}  from quantum chemistry. Using this method one can identify potential static correlation features of the obtained DFT ground states, which might call for more sophisticated multi-reference methods. To this end, we carry out calculation on the same 261 atom cluster model, Fig.~\ref{fig:structure_fod}(a), using  PBE functional and def2-SVP basis set. 

The FOD analysis performed on the ground state geometry reveals no  FOD density, see Fig.~\ref{fig:structure_fod}(b).
For the excited state geometry, however, some FOD density is predicted, which is found to be localized around the defect center on the s N-B bonds stretched along the symmetry axis, see Fig.~\ref{fig:structure_fod}(c). The shape of the FOD surface resembles the union of the two HOMOs and the LUMO with no detectable contribution from other orbitals. Thus, static correlation is only expected to be present in the aforementioned 4-electron 3-orbital (4,3) active space. As CASSCF(4,3) calculation did not result in mixed states, it can be concluded that the obtained FOD plot indicates merely the quasi-degeneracy of the frontier orbitals and does not stem from actual multireference character.  
Accordingly, we continue our analysis using TDDFT and constrained-DFT methods.

The PBE-TDDFT result obtained in a 261-atom cluster using def2-SVP basis set and the PBE constrained DFT result obtained in a 768-atom periodic model are provided in Table~\ref{tab:ZPL}.  As can be seen, the two methods agree very well with each other indicating a proper basis set selection for the cluster model. On the other hand, PBE-DFT and the NEVPT2 results deviates by $\approx$0.8~eV. Since semi-local exchange-correlation functionals tend to underestimate and the NEVPT2 ($\sim$MP2) tends to overestimate excitation energies, 
we set our focus to hybrid DFT methods, which can be grasped as an interpolation between  (semi-)local density functional theory and wavefunction theory.

In order to further narrow down the set of applicable methods, 
we select the HSE06\cite{HSE06} hybrid-DFT functional 
for further investigations; nevertheless, we varied the mixing parameter ($\alpha$, indicating the proportion of HF exchange in the functional formula) and thereby generated different HSE($\alpha$) methods. As the next step, we calculated the ZPL energy at HSE($\alpha$) levels of theory in 768-atom periodic supercell model. 

The most commonly used mixing parameter $\alpha$ range between 0.3 and 0.35 in hBN. These values are set by adjusting the theoretical band gap to the experimental one.\cite{weston_native_2018} While this method improves the description of host states and localized orbitals of intrinsic defects, it may not work for all defects, especially when stretched bonds are observed, see Fig.~\ref{fig:energylevels}(b). 

In order to test the accuracy of the functional for the description of a given state, we used the generalized Koopman's theorem\cite{lany_polaronic_2009,lany_generalized_2010, ivady_role_2013}  (ionization potential theory\cite{Perdew91,Perdew97,Almbladh85} in other terminology). According to this theorem, the eigenenergy of the highest occupied Kohn-Sham orbital and the ionization energy of the system should be equal when the exact exchange-correlation functional is used. Since the exact DFT functional is not known, approximate functionals may fail to fulfill this theorem. This failure manifests itself as a difference between the KS eigenenergy and the ionization energy. The non-Koopman's energy $E_{\text{NK}}$ measures this as
\begin{equation}
    E_{\text{NK}} = \varepsilon_{\text{HO}} - \left( E_{N+1} - E_{N} \right) \text{,}
\end{equation}
where $E_{N+1}$ and $E_{N}$ is the total energy of the system consisting of $N+1$ and $N$ electrons, and $\varepsilon_{\text{HO}}$ is the Kohn-Sham energy of the highest occupied orbital in the $N+1$-electron system. In practice, the non-Koopman's energy calculated in periodic boundary conditions depends also on technical parameters of the utilized model that needs to be taken into account,
\begin{equation}\label{eq:finENK}
    E_{\text{NK}} = E_{\text{NK}}^{\prime} \! \left( q,  L_{\perp}, L_{\parallel} \right) + \epsilon_{\text{corr}} \! \left( q, L_{\perp}, L_{\parallel} \right),
\end{equation}
where $E_{\text{NK}}^{\prime}$ is the non-Koopman's energy calculated in a periodic supercell model and  $\epsilon_{\text{corr}}$ is a correction term that cancels finite-size effect of the energy term originating mainly from spurious electrostatic interaction in periodic boundary conditions. Since the initial $N+1$ electron system is the negative charge state of the N$_i$ defect, $\epsilon_{\text{corr}}$ incorporates the charge correction of both $\varepsilon_{\text{HO}}$ and $E_{N+1}$. Both $E_{\text{NK}}^{\prime}$ and the $\epsilon_{\text{corr}}$ energy terms depend on the perpendicular and parallel to $c$ axis extension of the supercell,  $L_{\perp}$ and $L_{\parallel}$ respectively, as well as the charge state $q$ of the defect.

The non-Koopman's energy defined in Eqs.~(1) and (2) is a good indicator of the accuracy of DFT functionals for localized defect states. When a functional with tuneable parameters is used, such as the HSE($\alpha$) functional, adjustment of the free parameter may be utilized to decrease the non-Koopman's energy and thus to improve the description of the defect orbitals. This strategy has been successfully employed for several other point defect systems, see for instance Ref.~[\onlinecite{ivady_role_2013}]. On the other hand, we must first calculate the $\epsilon_{\text{corr}}$ term, which may be comparable in absolute value with the non-Koopman's energy itself. To proceed with the analysis on the adiabatic excited state-ground state energy difference of the N$_i$(-) defect, we carefully investigate the value of the $\epsilon_{\text{corr}}$ charge correction term.

\subsubsection{Charge correction of the N$_i(-)$ defect in hBN}

Charge correction of layers and surfaces is an active field of research today, however, the literature on charge correction methods of bulk models is fairly established.\cite{Freysoldt,lany_assessment_2008} The most accurate, but computationally expensive approach is to carry out a finite-size scaling test and extrapolate energy terms to the single defect limit. To quantify  $\epsilon_{\text{corr}}$ in Eq.~(2), we apply this procedure and consider 11 supercells of different sizes, each of which includes a single N$_i$ defect either in the negative ($N+1$ electron system) or in the neutral charge state ($N$ electron system). Due to the large computational demand of finite-size scaling tests, we carry out this study using PBE functional. We further assume that the results of finite-size scaling are independent of the choice of the exchange-correlation functional.

\begin{figure*}[!h]
\includegraphics[width=0.55\textwidth]{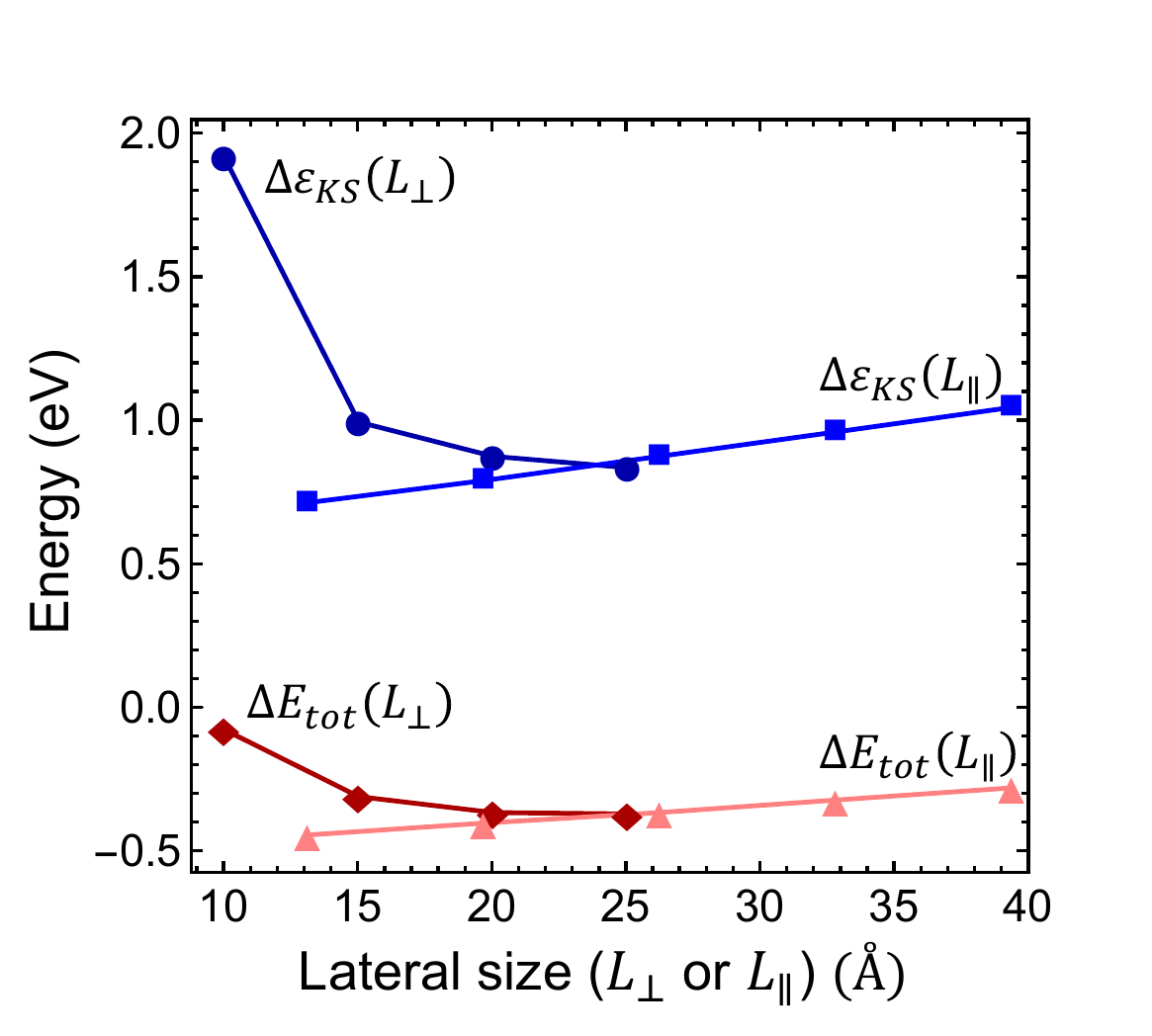}
\caption{ Finite-size scaling of the Kohn-Sham energy and the total energy difference. $\Delta\varepsilon_{KS}$ is the Kohn-Sham energy of the highest occupied defect orbital measured from the energy of the valance band maximum of the pristine supercell and $\Delta E_{tot} = E_{N+1}-{E_N}$ is the ionization energy of the negatively charged N$_i$ defect. Dark blue line with circles and dark red line with diamonds show the scaling of the energies as a function of the in-plane supercell size $L_{\perp}$, while light blue line with squares and pink line with triangles show the energies as a function of the perpendicular to plane supercell size $L_{\parallel}$.}
\label{fig:scale-1}
\end{figure*}

Interestingly, the scaling of the Kohn-Sham energy and the total energy shows an unexpected behavior as depicted in Fig.~\ref{fig:scale-1}. Considering the $L_{\perp}$ perpendicular to $c$ lateral size dependence of both the Kohn-Sham energy and the ionization energy, we observe converging curves that take finite values at the $L_{\perp} \rightarrow \infty$ limit. This behaviour is typical for bulk models. On the other hand, for the $L_{\parallel}$ parallel to $c$ lateral size of the supercell, we obtain a linear dependence, see Fig.~\ref{fig:scale-1}, that diverges at the $L_{\parallel} \rightarrow \infty$ limit. This odd behaviour is unexpected for bulk model of several layers of hBN sheets in periodic boundary conditions. Similar linear dependence is, however, observed in single layer hBN sheet in vacuum in periodic boundary conditions\cite{wang_determination_2015,komsa_charged_2014}, where the electrostatic interaction of the infinite charged layers diverges. Therefore, we conclude that the charge correction of Kohn-Sham and total energy terms in bulk hBN model cannot be described by charge correction methods developed for conventional bulk semiconductors, such as diamond and silicon. The charge correction should account for the electrostatics of a localized charge in a 2D layer immersed in the media of relative permittivity  other than 1 (due to the presence of other pristine hBN layers).

\begin{figure*}[!h]
\includegraphics[width=0.99\textwidth]{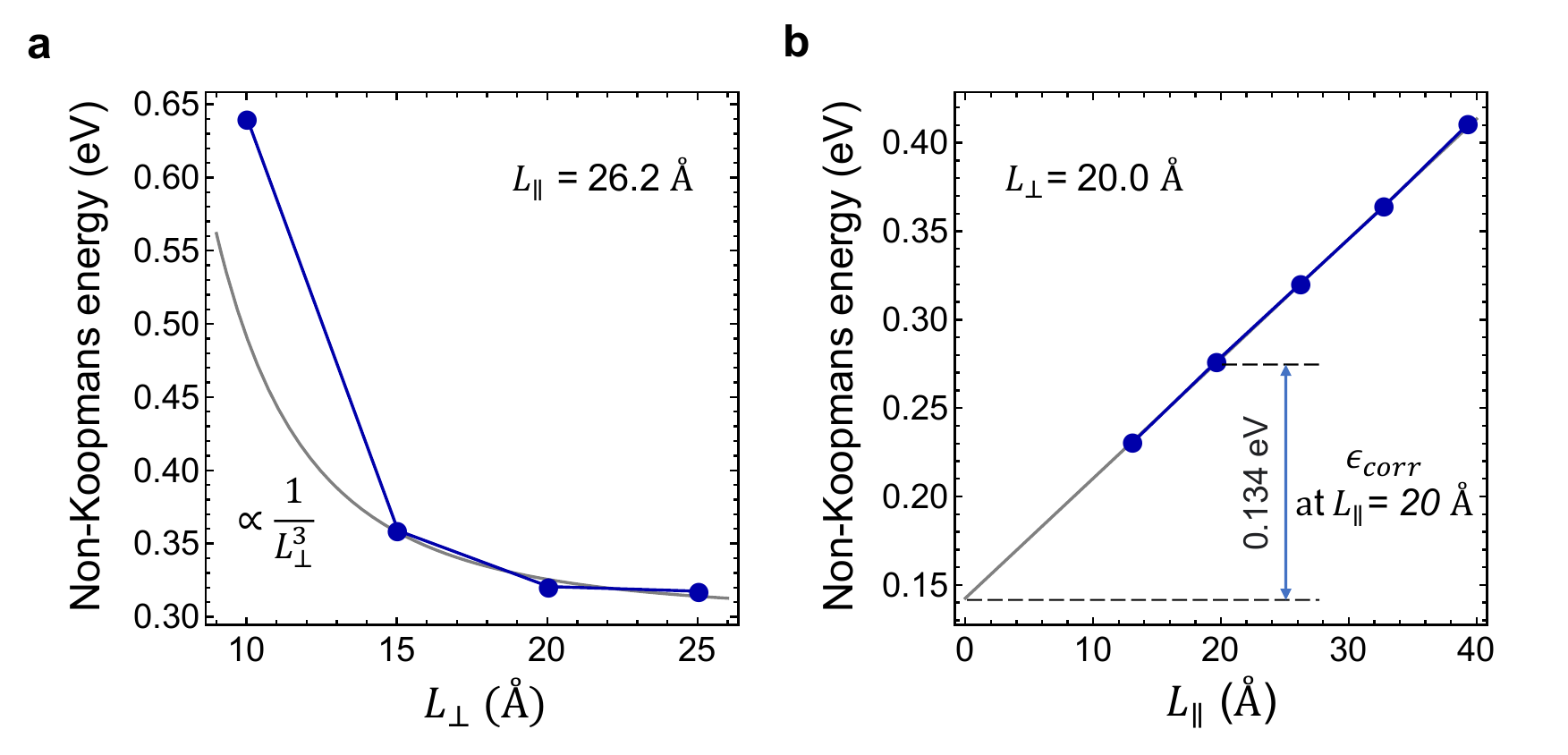}
\caption{ Determination of the charge correction of the Non-Koopmans energy. \textbf{a} and \textbf{b} depict the in-plane ($L_{\perp}$) and the parallel to $c$ ($L_{\parallel}$) supercell size dependence of the Non-Koopman's energy ($E_{NK}^{\prime}$). $\epsilon_{\text{corr}}$ is the charge correction of the Non-Koopman's energy at $L_{\parallel} = 20$~\AA . }
\label{fig:scale-2}
\end{figure*}

In order to model the systems, we express finite-size and charge dependent energy terms as\cite{wang_determination_2015}
\begin{equation} \label{eq:sizedep}
    \mathcal{E} \! \left( q,  L_{\perp}, L_{\parallel} \right) = \mathcal{E}_0 + \frac{a}{L_{\perp}} + \frac{b}{L_{\perp}^3} + c \frac{L_{\parallel}}{S},
\end{equation}
where $\mathcal{E}$ refers to either a Kohn-sham energy or the total energy of the system, $\mathcal{E}_0$ is the finite-size effect-free energy value that corresponds to the limit of a single point defect in the material, $a$, $b$, and $c$ are the free parameters that we will use to fit Eq.~(3) to the calculated energy curves. $S$ is the surface of the hBN layers in our finite periodic models. Note that for infinite layers, i.e. $S \rightarrow \infty$, the last term on the right hand side of Eq.~(\ref{eq:sizedep}) vanishes. For finite supercells the last term is of high relevance, though.   Since we are interested in the non-Koopman's energy, we calculate the supercell size dependence of this energy term in the ground state and fit Eq.~(3) directly to the obtained curve, see Fig.~\ref{fig:scale-2}. 

Considering the perpendicular to $c$ lateral size dependence of the non-Koopman's energy at $L_{\parallel} = 26.2$~\AA , we observe converging tendency, see Fig.~\ref{fig:scale-2}(a), which is best fitted by a $1/L_{\perp}^3$ function using points beyond 15~\AA . The smallest supercell is apparently an outlier of the general trend expectedly due to the overlap of the defect states. 
As can be seen in Fig.~\ref{fig:scale-2}(a), the non-Koopman's energy only slightly changes beyond $L_{\perp} = 20$~\AA , thus we consider supercells of this later size to be convergent in the perpendicular to $c$ direction. Therefore, for $L_{\perp} \geq 20$~\AA\  we set parameters $a$ and $b$ to zero in Eq.~(3), which approximation causes an error no larger than 15~meV. Parallel to $c$ supercell size dependence of the non-Koopman's energy for $L_{\perp} = 20.0$~\AA\ is depicted in Fig.~\ref{fig:scale-2}(b). The points can be perfectly fitted by a linear curve, which intersects the energy axis at 0.145~eV for $L_{\parallel} = 0$. From the extrapolation to zero, we can obtain the finite-size effect free non-Koompan's energy for the considered charged layered system, see Ref.~\onlinecite{wang_determination_2015}. For a 20~\AA\ $\times$ 20~\AA\ $\times$ 20~\AA\ supercell consisting of 768 atoms, the spurious electrostatic energy correction to the non-Koopman's energy, see Eq.~(\ref{eq:finENK}), is $\varepsilon_{\text{corr}} = E_{\text{NK}} - E_{\text{NK}}^{\prime} \! \left( -1,  20.0, 20.0 \right) = 0.134$~eV. The periodic model related electrostatic interaction between the negatively charged defect and the homogeneous positive background depends negligibly on the small variation of the defect states due to the choice of the functional. Therefore, we use the same correction for different HSE($\alpha$) functionals in the same supercell of 768 atoms. 

\begin{figure*}[!h]
\includegraphics[width=0.55\textwidth]{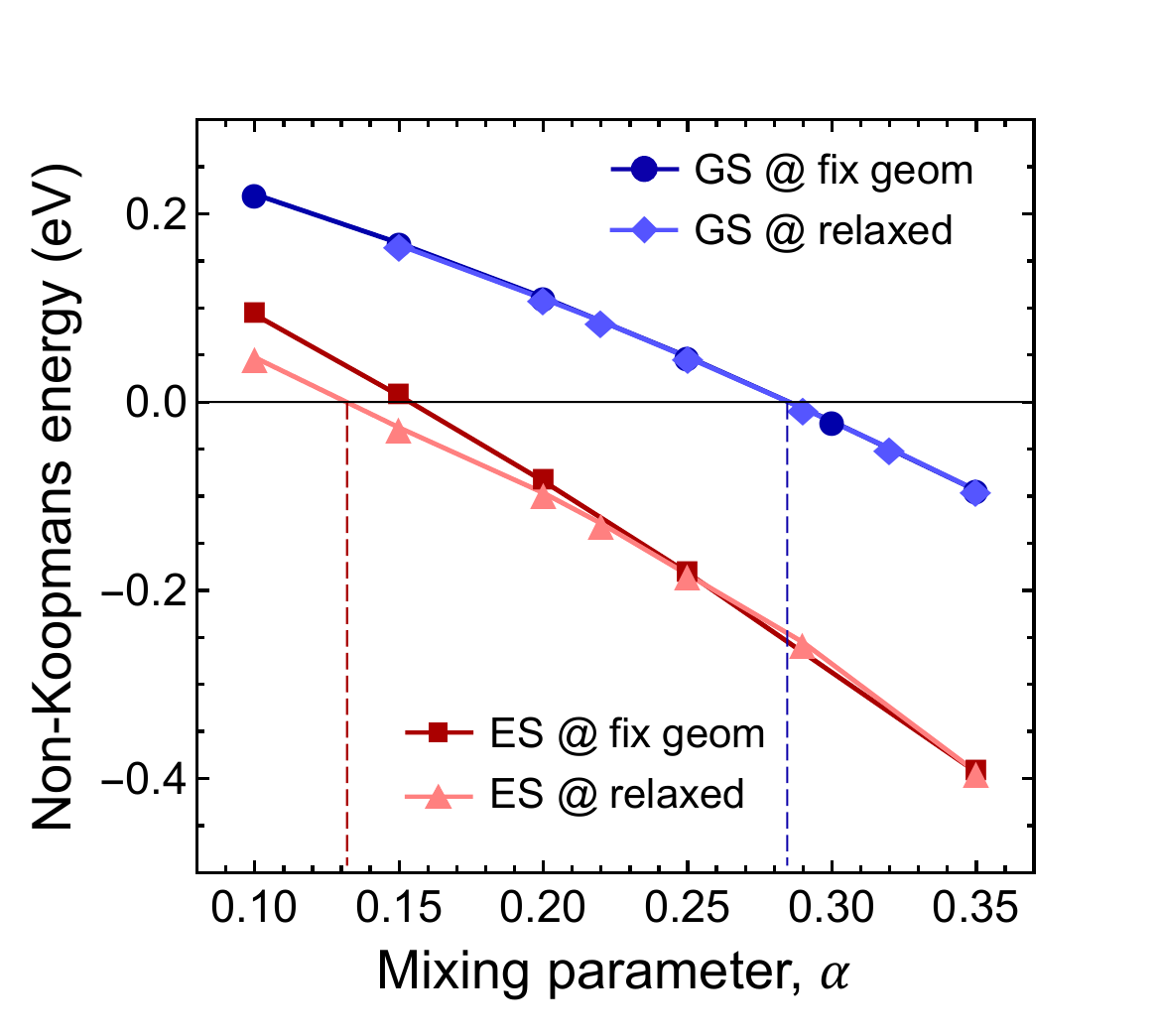}
\caption{ Mixing parameter dependence of the finite-size effect corrected non-Koompan's energy. Dark blue line with circles and dark red line with squares depict the non-Koopman's energy ($E_{\text{NK}}$) calculated for the highest occupied defect state in the ground and the excited states on fixed geometries obtained with HSE(0.3) functional. Light blue line with diamonds and pink line with triangles show the mixing parameter dependence of the Koopman's energy in the ground and excited states obtained on corresponding relaxed geometries. }
\label{fig:alpha}
\end{figure*}

\subsubsection{Optimization of the HSE($\alpha$) functional for the electronic structure of the N$_i$(-) defect}

Having the charge correction of the non-Koopman's energy determined, we compute the finite-size effect free non-Koopman's energy value according to Eq.~(\ref{eq:finENK}). In order to evaluate the accuracy of HSE($\alpha$) functional, we calculate the mixing parameter $\alpha$ dependence of the non-Koopman's energy for the ground and the excited states with and without structural relaxation, see Fig.~\ref{fig:alpha}. When no relaxation is taken into account, we use optimized ground and excited state structures as obtained by the HSE(0.30) functional.

First, we study the accuracy of the HSE($\alpha$) functional on the ground state of the negatively charged N$_i$ defect. As can be seen in Fig.~\ref{fig:alpha}, $E_{\text{NK}} = 0 $ is achieved at $\alpha = 0.285$. Note that geometry relaxation has only negligible effects on the non-Koopman's energy, see Fig.~\ref{fig:alpha}. The obtained optimal mixing parameter is close to $\alpha = 0.3$ used in the analysis of the ground state properties, thus our results presented for the stability of the N$_i$ defect are valid. On the other hand, for the excited state we observe a different behaviour. At $\alpha > 0.3$ the finite-size effect corrected non-Koopman's energy takes a significant negative value, indicating that the frequently used HSE(\{0.3,0.32,0.35\}) functionals are not appropriate for the description of the excited state of the N$_i$(-) defect.

For the excited state, the non-Koopman's energy vanishes at $\alpha = 0.132$ ($\alpha \approx 0.154$) when structural relaxation is taken (not taken) into account. Here, structural relaxation makes a difference in the non-Koopman's energy in contrast to the case of the ground state. These results clearly indicate the need for the reduction of the exact exchange contribution in the excited state. Since accurate description of the ground state and the excited state requires two different functionals, HSE(0.132) and HSE(0.285), the adiabatic energy differences of the states cannot be calculated with the highest accuracy. On the other hand, we can use our results to significantly narrow down the uncertainly of the theoretical predictions and provide the most likely value of the adiabatic energy difference and ZPL. 

\begin{figure*}[!h]
\includegraphics[width=0.55\textwidth]{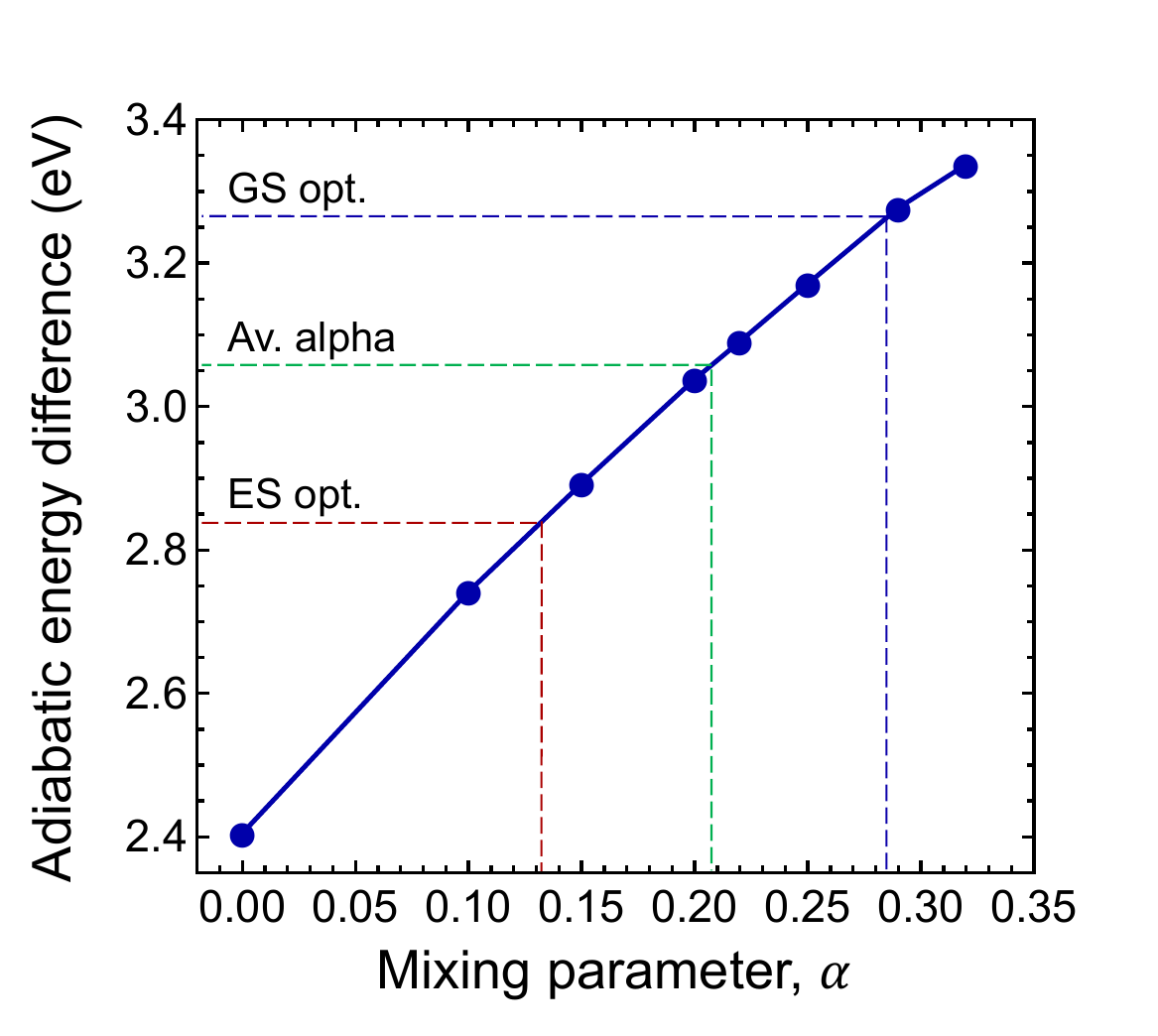}
\caption{ Mixing parameter dependence of the adiabatic energy difference of the ground and excited states ($E_{\text{AD}}$). }
\label{fig:zpl-alpha}
\end{figure*}

To this end, we calculate the mixing parameter dependence of the adiabatic energy difference of the ground and excited states, see Fig.~\ref{fig:zpl-alpha} and Table~\ref{tab:ZPL}. As can be seen, the energy difference decreases as the exact exchange contribution is reduced. At $\alpha = 0.285$ and $\alpha = 0.132$ the ground state and the excited state is described with high accuracy, respectively. The adiabatic energy differences corresponding to these mixing parameter values are 3.27~eV and 2.84~eV. We assume that the true value of the adiabatic energy difference could be found in this interval. On the other hand, due to the different behaviour of the ground and excited states, the accurate value of the adiabatic energy difference cannot be obtained by a HSE($\alpha$) functional using a single value of the mixing parameter. At the same time energy differences cannot be computed using two different funtionals. Therefore, we define the most plausible value of the adiabatic energy difference by the mean of the values obtained with the ground and excited state optimized functionals, which is $E_{\text{AD}} = 3.06$~eV. Due to the approximately linear dependence of the adiabatic energy difference on the mixing parameter, the same value can be obtained by HSE($\overline{\alpha}$), where $\overline{\alpha}$ is the mean value of the optimized mixing parameters. The expected error margin of the adiabatic energy difference is approximately $\pm0.2$~eV.

\subsection{Electron-phonon coupling}

Next, we investigate local vibrations and the phonon side band of the photoluminescence spectrum of the negatively charged nitrogen split interstitial defect in hBN. For this purpose, we utilize cluster models of various sizes. All DFT calculations on the cluster models are performed using the def2-SVP basis~\cite{weigend2005balanced} and PBE exchange correlation functional~\cite{perdew1996generalized}. In order to preserve the distance of the van der Waals layers in line with experimental expectations, only the chemically most relevant atoms are considered in the geometry relaxation procedure. We demonstrate that the cluster model is capable of providing convergent predictions for the studied system  by calculating the adiabatic energy difference of the ground and excited states  with varying flake size, number of stacked layers, and geometrical  optimization scheme. For further discussion see Appendix A.

For the PL spectrum calculations, three different cluster models are constructed from the bulk geometry relaxed by VASP on PBE level: i) a 3 layer model with relatively large layer size (L),  ii) a 3 layer and iii) a 5 layer model with smaller layer size (S), see Fig.~\ref{fig:structure_fod} and Appendix A. The atomic clusters are capped with hydrogen atoms at the edges leading to B$_{110}$N$_{110}$H$_{63}$, B$_{56}$N$_{56}$H$_{46}$, and B$_{92}$N$_{94}$H$_{75}$ clusters, respectively.

In order to preserve the well defined layered structure of the standalone cluster, its geometry is only partially relaxed on PBE level of theory by ORCA. 
In a minimalist case, besides the terminating hydrogens, only the four atoms in the center of each layer and the interstitial N atom are considered in geometry optimization, see atoms highlighted in translucent red in Fig.~\ref{fig:structure_fod}a. Since in this scheme we only optimize a small set of atoms,  we refer to this relaxation strategy as S in the following discussion.
In the other case, this subset is further expanded by the outer atoms of the 3 hexagonal ring around the center of each layer,  see atoms highlighted in red and yellow in Fig.~\ref{fig:structure_fod}a. In this case  the position of 13 second-row atoms per layer near the center of the molecule plus the interstitial N are relaxed. 
Minimizing a medium set of atoms, we refer it to as strategy M.
In the third case, denoted strategy L, the further shell of outer atoms are also optimized, highlighted in green in Fig.~\ref{fig:structure_fod}a, i.e., 37 atoms per layer plus the interstitial N are optimized.
The various setups are labeled as A-B-C-D, where A is the number of layers in the cluster, B is the number of optimized layers, C is the size of the layers, and D refers to the region which was optimized in the layers. For example, 5-3-S-M means that a 5 layer model with the smaller layer was utilized and only 13 atoms per layer was optimized in the 3 middle layers plus the interstitial N.

\begin{figure}[!t]
\centering
\includegraphics[width=0.45\textwidth]{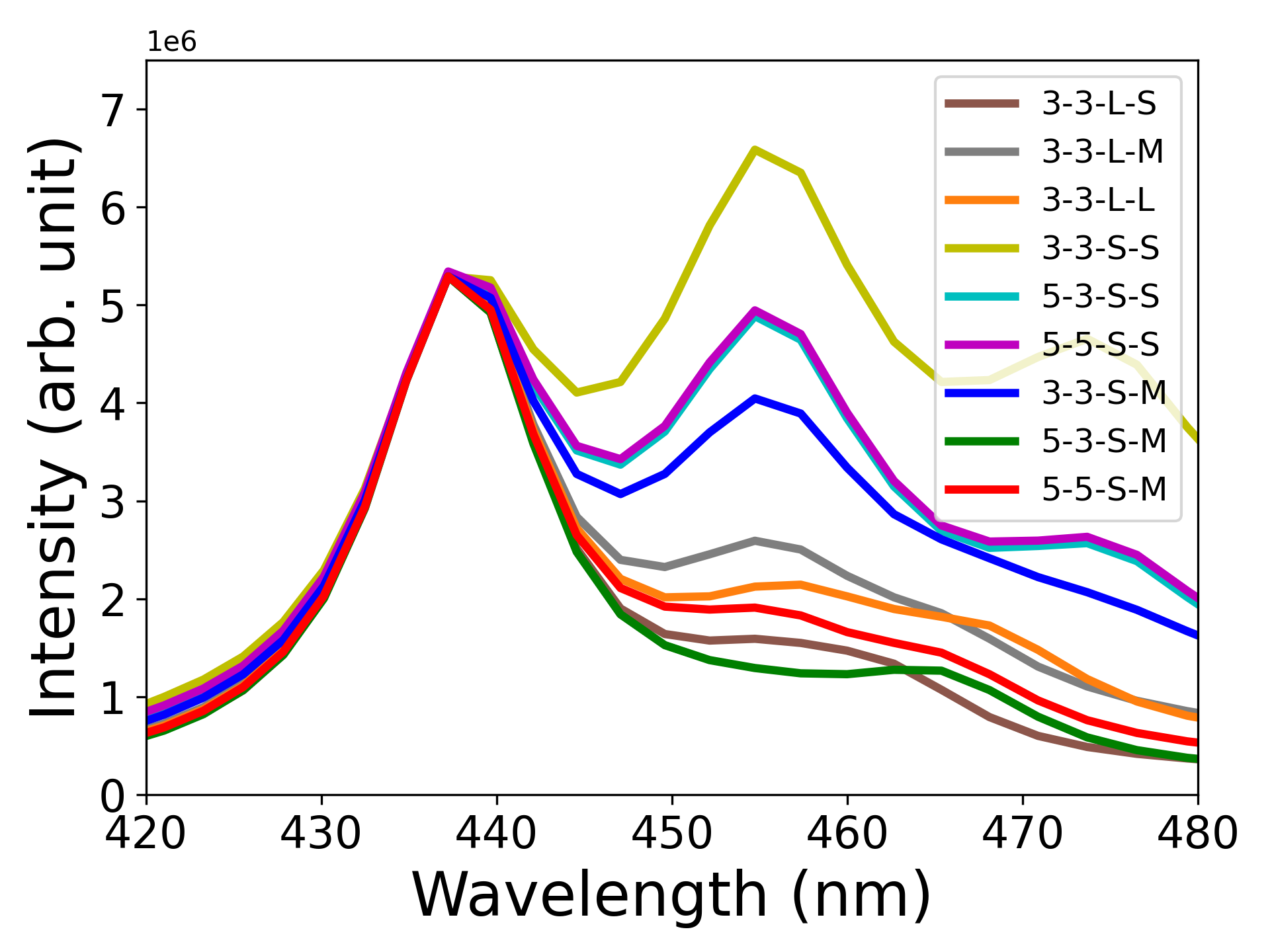}
\includegraphics[width=0.45\textwidth]{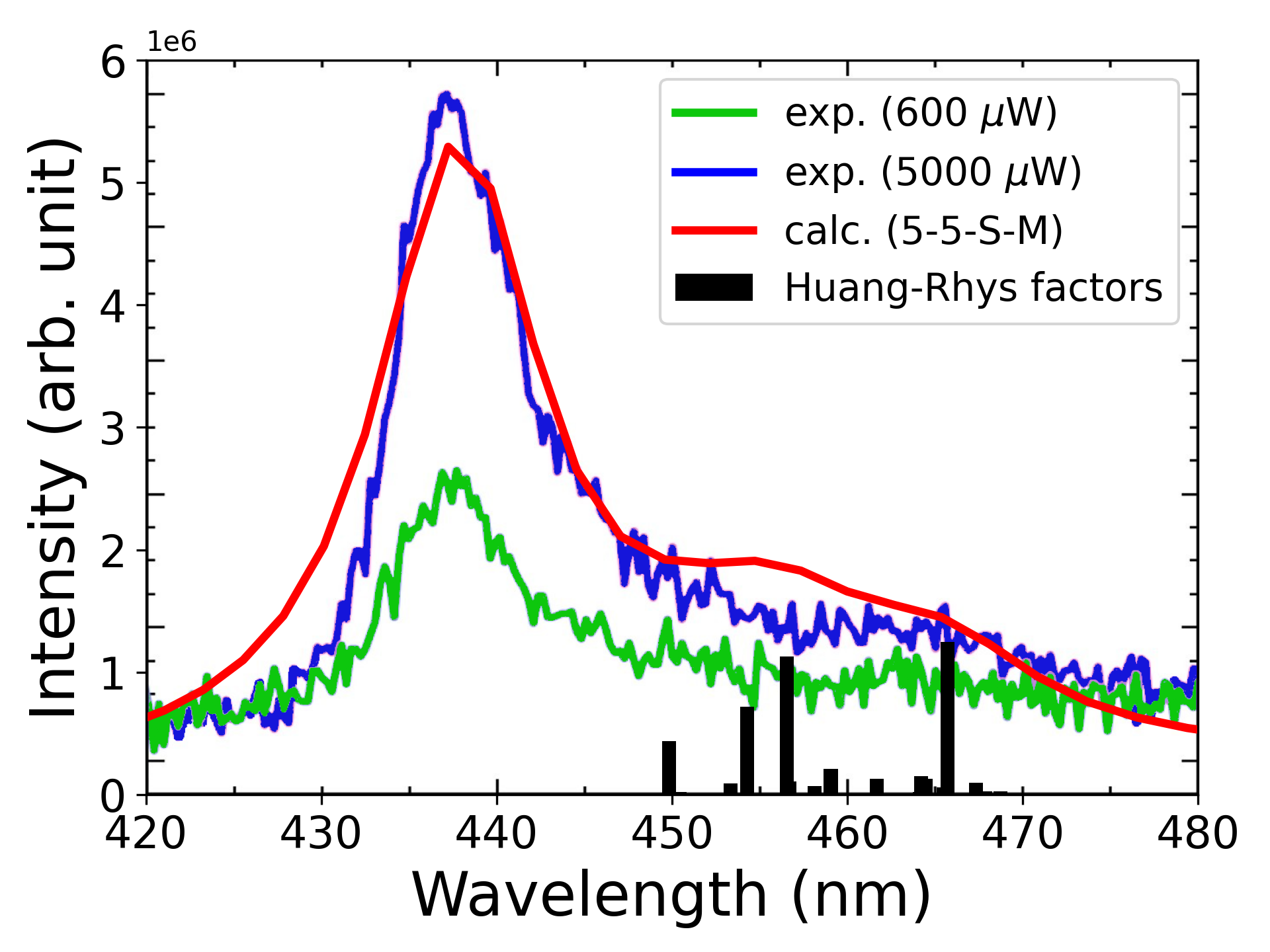}
\caption{\label{fig:pl} 
 Left: calculated PL spectra with different geometry optimization setups. Right: The best calculated PL spectrum compared to experimental results\cite{zhigulin_photophysics_2023}. Partial Huang-Rhys factors of each mode are also presented with bars. Linewidth is set to 300 cm$^{-1}$. For better comparison of the theoretical and the experimental PL spectra, the theoretical spectrum is shifted to match the experimental ZPL.} 
\end{figure}

\begin{figure}[!t]
\centering
\begin{tabular}{cc}
 \includegraphics[width=0.3\textwidth]{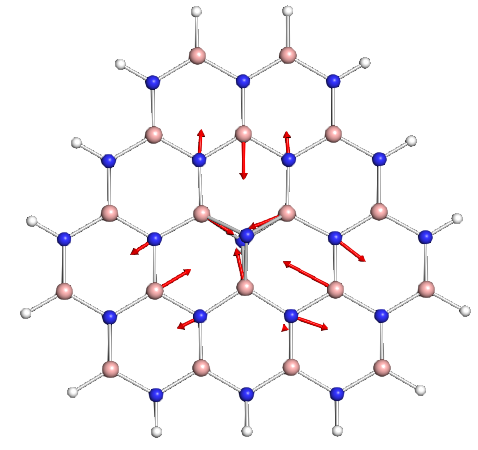}    & 
  \includegraphics[width=0.45\textwidth]{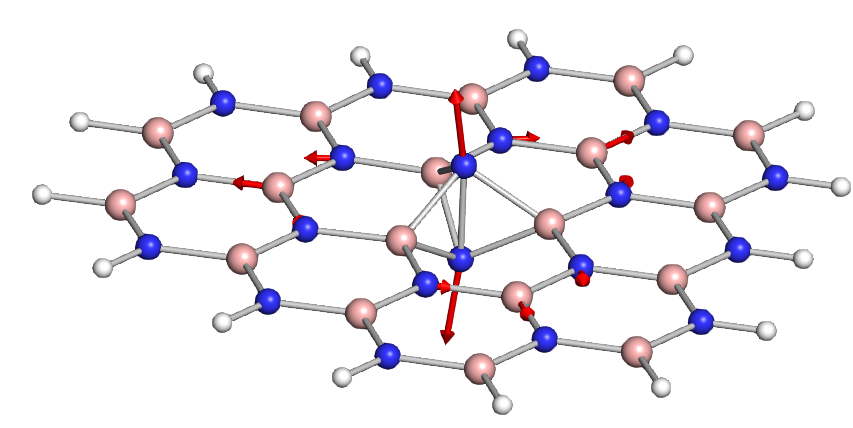} \\
 frequency: 1358.54 cm$^-1$ & frequency: 928.15 cm$^-1$ \\
 Huang-Rhys factor: 0.125 & Huang-Rhys factor: 0.113\\
    \includegraphics[width=0.45\textwidth]{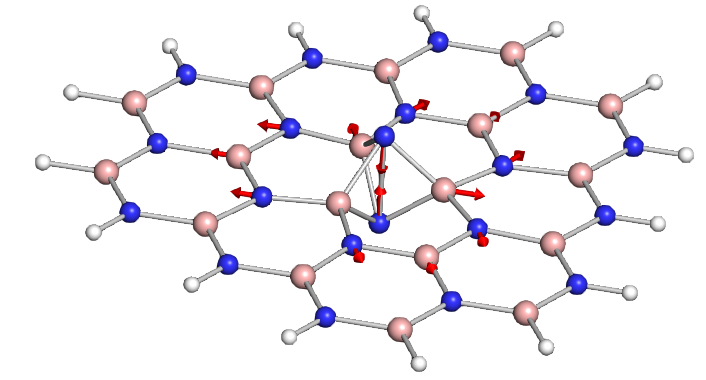}   & 
      \includegraphics[width=0.45\textwidth]{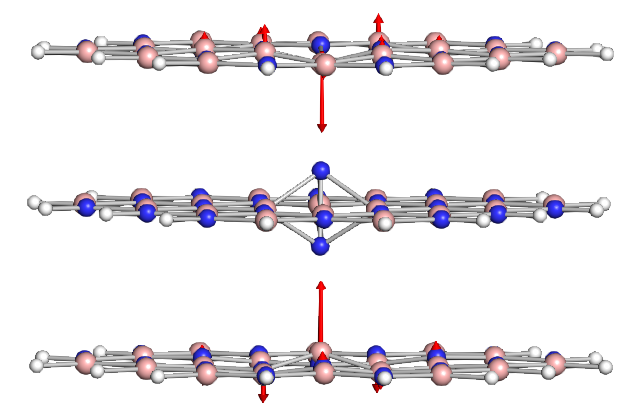} \\
  frequency: 818.20 cm$^-1$ & frequency: 600.69 cm$^-1$ \\
Huang-Rhys factor: 0.072 & Huang-Rhys factor: 0.044\\
\end{tabular}
\caption{\label{fig:modes} 
Visualization of the four most dominant phonon modes based on the Huang-Rhys factors.
Note that these interesting modes are localized around the defect center.
} 
\end{figure}

Convergence of the photoluminescence spectra as a function of the cluster size and relaxation strategy  is depicted in Fig.~\ref{fig:pl}a. Due to the relatively large change of the position of the nitrogen atoms and the first neighbor atoms along the symmetry axis, see Fig.~\ref{fig:energylevels}, the vertical gradient approximation\cite{petrenko2012efficient,ferrer2012comparison} is utilized. 
In case of the smallest model of three layers (3-3-S-S), the intensity of the second peak is larger compared to the first peak associated to the ZPL. 
Similarly, the 3 layer model with the M region optimized (3-3-S-M) exhibits a large sideband.
On the contrary, the two spectra from the largest 5 layer models (5-3-S-M and 5-5-S-M) show a strong ZPL emission and much less pronounced sideband. The same can be said if the larger layer is used (L) which implies that convergence of the PL spectrum can be reached in the flake models. 
Note that in Fig.~\ref{fig:pl}a all transition lines, including the ZPL, were broadened by an empirical 300~cm$^{-1}$ linewidth.

For the convergent model (5-5-S-M) we obtain a Huang-Rhys factor of 0.491, and radiative lifetime of 54 ns using a refractive index equal to 2.13. The corresponding transition dipole moment vector is (0.37574,-0.19576,-0.00215) in atomic unit, where the $z$ direction is parallel to the $c$ axis of hBN.  It must be noted that due to the large difference between the ground and excited state, our investigation is limited to using the vertical gradient approximation that only uses the ground state geometry and Hessian. The transition dipole moment between the two states could differ considerably from what we obtain from PBE-TDDFT on the excited state geometry. In principle, we expect the lifetime to be overestimated in our calculations. 

From the analysis of the partial Huang-Rhys factors, we deduce that the dominant local vibrations responsible for the phonon sideband are
i) the movement of the atoms surrounding the defect causing the Jahn-Teller distortion, 
ii) the stretching and contraction of the N-N bond in the defect, and
iii) the movement of the B atoms from the neighbouring layers towards the defect, see Fig.~\ref{fig:modes}.
These vibrations account for the geometry difference between the ground and excited states.

Lastly, we calculate the zero-point energy (ZPE) contribution to the adiabatic energy difference obtained in the previous section. From the sum of the two contributions we compute the ZPL energy.  The ZPE contribution is approximated by numerical frequency calculations where a 5 layer model with 261 atom is used and the middle 13 atoms are relaxed in each layer plus the interstitial nitrogen. For the ZPE contribution we obtain -0.10~eV meaning that the zero point energy contribution of the local vibrational modes is larger in the ground state than in the excited state. This is understandable as stretched bonds are formed in the excited state configuration that give rise a softer potential and lower vibrational frequencies. We assume in this paper that the zero point energy contribution does not significantly depend on the electronic structure calculation method used, thus we use the same zero point energy contribution in all ZPL value approximations. The third row of Table~\ref{tab:ZPL} provides the ZPE corrected adiabatic energy differences as obtained with different methods.

\subsection{Electric field dependence of the ZPL energy}

Finally, we numerically investigate the electric field dependence of the of the ZPL line of the negatively charged nitrogen split interstitial in hBN. As discussed in Ref.~[\onlinecite{zhigulin_stark_2022,zhigulin_photophysics_2023}], defects exhibiting high D$_{3h}$ symmetry in both the ground and excited state may be less sensitive to external electric fields and only second or higher order electric field dependence is expected. The negatively charged N$_{i}$ defect does exhibit  D$_{3h}$ symmetry in the ground state, as also discussed in Ref.~[\onlinecite{zhigulin_stark_2022}]; however, the excited state is Jahn-Teller unstable that lowers the symmetry to C$_{\text{2v}}$. Therefore, it is a question whether the desirable quadratic dependence of the ZPL energy on the electric field is preserved even in such a low symmetry. Our preceding study\cite{zhigulin_stark_2022} suggests that the quadratic term dominates the electric field dependence of the ZPL, however, due to the applied large-scale model and consequent numerical uncertainties the results need to be reconsidered by an improved independent model. Here, we apply a cluster model, atomic basis functions, and ORCA suit opposed to the slab model, plane wave basis set, and VASP package utilized previously\cite{zhigulin_stark_2022}.

To study the effect of the electric field on the zero-phonon line  of the N$_i$(-) defect in hBN, a static field is applied perpendicular and parallel to the layers of the cluster model with varying strength ranging from -0.05 to 0.05 V/\AA{}. 
For the calculations, we use PBE-TDDFT method, which somewhat underestimates the ZPL energy, see Table~\ref{tab:ZPL}, however, it can capture to the polarization of the defect orbitals induced by the external electric field. We expect our results to be qualitatively accurate.

\begin{figure*}[!t]
\includegraphics[width=0.55\textwidth]{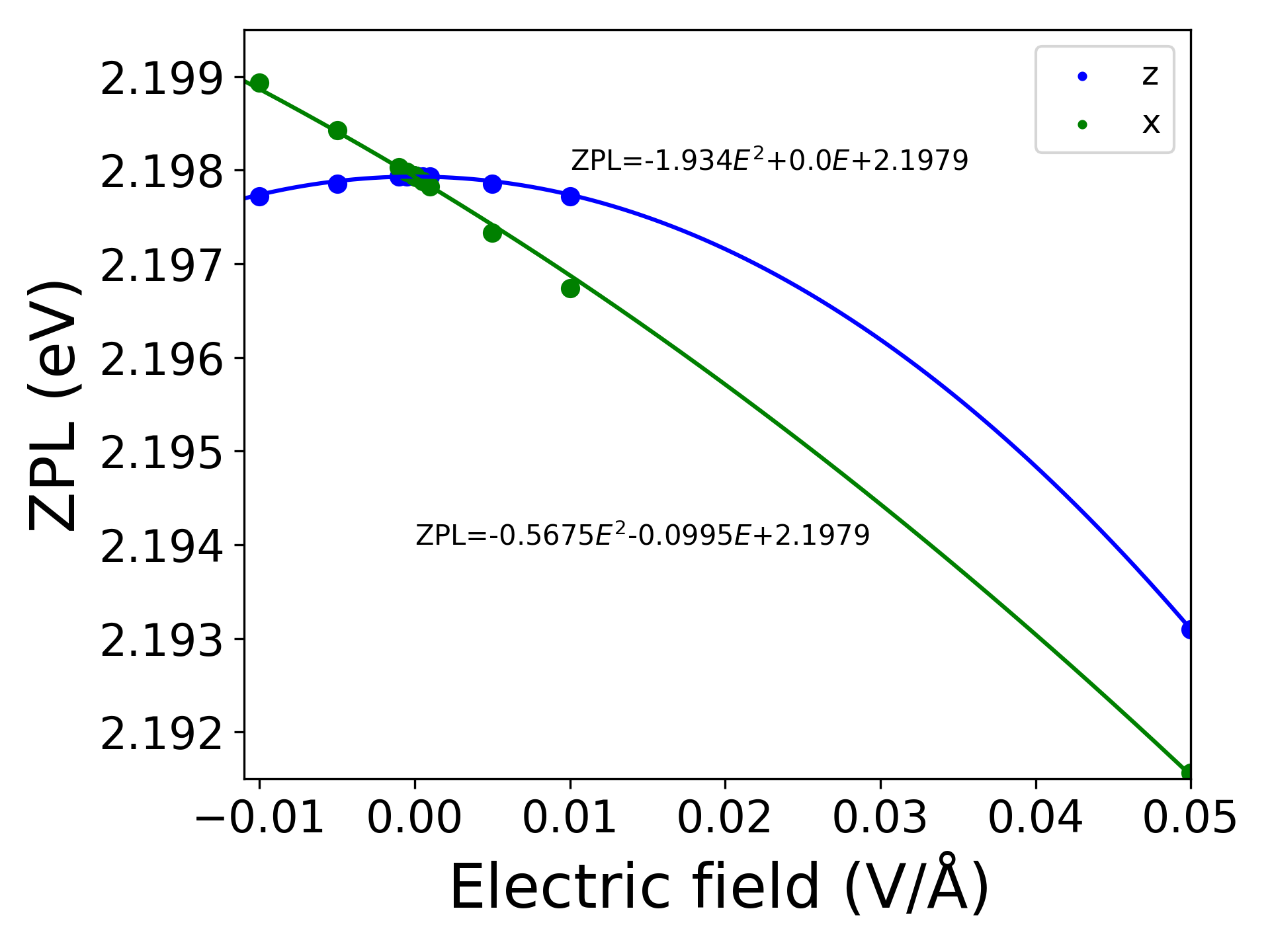}
\caption{\label{fig:ZPLvsEfield} The electric field dependence of the ZPL. The ZPL energy is calculated on PBE-TDDFT level of theory in a 261-atom cluster model. Due to the approximations used, the absolute value of the ZPL energy is underestimated.
}
\end{figure*}

As can be seen in Fig.~\ref{fig:ZPLvsEfield}, perpendicular to the layers we find that the ZPL energy decreases quadratically with increasing strength of electric field. However, if the electric field is parallel to the layers, the ZPL energy depends almost linearly on the strength of the electric field. 
This behaviour is expected from the changes in dipole moment ($\Delta \mu = (-0.67, 0.42, -0.01)$~Debye) and polarizibility ($\Delta \alpha$) in both directions. In the parallel to layer direction we obtain 55~$\AA{}^3$ for $\Delta \alpha$, and 0.01~Debye for $\Delta \mu$, while in perpendicular direction 17~$\AA{}^3$ and 0.67~Debye, respectively. We do not find a perfectly linear curve in $x$ direction due to the non-zero quadratic contribution to the Stark shift.

\section{Discussion - Plausible model for the blue emitter in hBN}

Finally, we compare our numerical results on the negatively charged nitrogen [0001] split interstitial to the experimental reports of the blue emitter in hBN\cite{zhigulin_photophysics_2023}. In summary, we propose the N$_i$(-) defect as a working model for the blue emitter, although not all the experimental observation can be explained with the current model. Below, we list all the known pros and cons of this assignment.

\bigbreak
\noindent{\emph{Pros}}
\bigbreak

The blue emitter in hBN can be intentionally created by electron irradiation. As a primary damage of high energy electron bombarding, boron and nitrogen vacancies and interstitials are created. Complex defect formation is expected as a result of secondary processes, such as migration of the created defects and possible recombination with other defects in the lattice. The nitrogen split interstitial is a primary radiation damage and it is expected to be formed in substantial concentrations. 

Regarding the optical properties of the blue emitter, the underlying microscopic structure is expected to possess D$_{3h}$ symmetry, while the relevant electronic structure of the defect should consists of a fully occupied defect state in the lower half of the band gap and an unoccupied defect state in the upper half of the band gap. The latter conditions are deduced from the stability of the defect under continuous excitation. The negatively charged nitrogen split interstitial defect possesses a fully occupied $e^{\prime\prime}$ state and an empty $a_2^{\prime\prime}$ state inside the band gap giving rise to a singlet ground state. This electronic structure is in agreement with the expectations derived from the observed properties of the blue emitter.

The blue emitter is known to emit light mostly along out of plane directions with in-plane electric field polarization. Interestingly, the perpendicular to $c$ polarization of the blue emitter is not homogeneous, there is a preferential in-plane polarisation direction.\cite{zhigulin_stark_2022} This observation also implies that the three-fold rotation symmetry of the defect should be violated either in the ground or the excited state, which has been overlooked previously. To explain this behaviour, we draw attention to the Jahn-Teller instability of the excited state of the N$_i$(-) defect. Since an electron is excited from the  $e^{\prime\prime}$ state to the $a_2^{\prime\prime}$ state, the symmetry is reduced to C$_{2v}$ in the excited state due to the Jahn-Teller distortion that gives rise to a clear single axis polarization pattern in the calculated transition dipole moment. This is in agreement with the observations.

The blue emitter exhibits a well resolved ZPL at 2.844~eV (436~nm)\cite{zhigulin_stark_2022}. While it is challenging to accurately calculate the ZPL energy of the nitrogen interstitial, our prediction agrees within the error margin of our calculations with the ZPL energy of the blue emitter. We obtained 2.96~eV (419~nm), which is 12~meV (17~nm)  blue shifted compared to the blue emitter's  ZPL. This deviation is well within the estimated error bar of $\pm0.2$~eV of our calculations.

Finally, we discuss the PL spectra of the blue emitter and the N$_i$(-) defect. As can be seen in Fig.~\ref{fig:pl}, the experimental and the convergent theoretical PL spectra agree very well with each other. Both spectra exhibit dominant emission in the ZPL and a vanishing sideband. For the N$_i$ defect we obtain a Huang-Rhys factor of 0.491 and Debye-Waller factor of 0.61.

Due to the Jahn-Teller distorted excited state, the N$_i$(-) defect possesses a non-zero electric dipole in this state. The dipole is in the plane of the hosting hBN layer. Depending on angle $\delta$ of in-plane electric field and the electric dipole of the defect, one may observe either linear ($\delta = \{0^\circ, 180^\circ\}$), or quadratic ($\delta = \{90^\circ, 270^\circ\}$), or mixed electric field dependence. This observation may explain the differences of the experimental results reported in Refs.~[\onlinecite{noh_stark_2018,nikolay_very_2019,zhigulin_stark_2022}]. 

\bigbreak
\noindent{\emph{Cons}}
\bigbreak

For perpendicular to plane electric fields we found  quadratic Stark shift, while the paper of Zhigulin et al in Ref.~[\onlinecite{zhigulin_stark_2022}] reports on a weak linear dependence. We attribute this difference to perpendicular to plane disturbances, for instance due to bending of the hBN sample. Such a distortion can lift the in-plane reflection symmetry of the system, which in turn enhances an out-of-plane electric dipole component.

The blue emitter possesses an excited state lifetime as short as $\sim 2.15$~ns.\cite{zhigulin_photophysics_2023} Our computational results suggest a longer radiative lifetime in the range of 54~ns. We note that the calculations account only for radiative processes, thus the experimental excited state is expectedly shorter than the theoretical estimates. However, the deviation between the two values cannot be only explained by nonradiative processes. We anticipate that the deviation is due to the approximations used in the calculations. 

Furthermore, according to recent study on the fabrication of the blue emitter, there seems to be a correlation between the appearance of the blue emitter and the UV emitter in hBN.\cite{gale_site-specific_2022,zhigulin_stark_2022,zhigulin_photophysics_2023} The latter has been assigned to the the carbon dimer recently.\cite{mackoit-sinkeviciene_carbon_2019} Since the nitrogen interstitial and the carbon dimer structures have no common roots, the connection between the two defects is not obvious. We anticipate that not the formation of the atomic structures are related, but instead the stability of the optical bright state is affected similar ways. For instance, due to their similar electronic structure, the position of the Fermi level may affect the two defects similarly. This may explain the presumed relationship of the appearance of the two defects.

\section{Methods}

\subsection{Further details on the calculation of periodic models using VASP}
 
The density functional theory calculations for bulk models are performed using Vienna Ab-initio simulation package (VASP)\cite{VASP2}, within the projector-augmented wave (PAW) method\cite{PAW} with a plane-wave cutoff of 500 eV. The exchange correlation is described by the generalized gradient approximation of Perdew, Burke, and Ernzerhof (PBE)\cite{perdew1996generalized}, and we also consider the screened hybrid functional of Heyd-Scuseria-Ernzerhof (HSE06)\cite{HSE06} with modified exact exchange fraction of 0.1-0.35 and screening parameter of 0.4. The van der Waals interaction is included using Grimme-D3 method with Becke-Johnson damping\cite{Grimme2010, Grimme2011}. 

The unit cell of bulk hBN is optimized with 15$\times$15$\times$5 k-point grid. We obtain lattice parameters $a=2.50$~\AA, $c=6.56$~\AA, which are in agreement with experimental values of $a=2.50$~\AA, $c=6.60$~\AA{} measured at 10 K\cite{BNlattice}. In the case of defect stability study, we consider orthorhombic supercell of 120 atoms  (5$a$, 3$a$+6$b$, $c$) as well as  hexagonal supercell containing 512 or 768 atoms. The reciprocal space is sampled with $\Gamma$-point. Atom relaxation is carried out until the forces are less than 0.01 eV/\AA\  (0.025 eV/\AA) with PBE (HSE($\alpha$)) functional. We also include the non-spherical contributions within the PAW spheres.

\subsection{Further details on the numerical studies of cluster models using ORCA}

The DFT calculations of the cluster models are made
by ORCA 5.0.3~\cite{neese2022software}. Most of the calculations are made with PBE functional and def2-SVP basis set, if otherwise not noted. During all calculation with ORCA resolution of identity (RI) approximation are used to the Coulomb term and the fitting basis are applied through the AutoAux feature.
Geometry optimizations are made with default settings.
For excited states different number of roots are required depending on the setups, and the corresponding root is chosen to be followed which represents the correct excited state. 
FOD analysis is performed at 5000 K as suggested in the seminal article\cite{grimme2015practicable}.
Numerical frequency calculations are performed on only those atoms which were originally optimized at the particular model setup with steps of 0.005 $\AA{}^3$. 
The PL spectra are determined based on the Hessians obtained from the numerical frequency calculations, using the vertical gradient approximation\cite{petrenko2012efficient,ferrer2012comparison} and a 300 cm$^-1$ linewidth. 

\section*{Appendix A}

In order to demonstrate that our cluster model is capable of providing convergent predictions for the studied system, we calculate the PBE-TDDFT ZPL energy  with varying flake size, number of stacked layers, and geometrical  optimization scheme. The corresponding numerical results are collected in Table ~\ref{table:size}. Note that ZPL energies in this study do not include ZPE contribution.

Interestingly, we find that the ground state is relatively prone to the relaxation, the energy gap increases by granting more freedom for geometry relaxation than the minimalist scheme S.
By adding external layers to the minimal model of three stacked layers,  the in-layer planarity of the excited state is preserved yielding some 0.10-0.2 eV energy gain.
This stabilization effect is found not to be sensitive to the relaxation of the additional covering layers.
In conclusion, independently of the investigated cluster setups, the lowest singlet excitation is consistently found at around 2-2.5~eV.
Based on the ZPL results in the following analysis we study the 5 layer model with smaller layers and optimize all of them with the M strategy.  
\begin{table}
\begin{tabular}{c|c|c|c||cc}
\multirow{2}{4em}{\# layers} & ~\# relaxed~ & ~layer~ & ~relaxation~ &  \multicolumn{2}{c}{ZPL (eV)} \\
& layers & size & scheme & PBE & HSE(0.25)\\
\hline
\hline
    3 & 3 & S & S & 2.005 & 3.030 \\
    3 & 3 & S & M & 2.398 & 3.359\\
\hline
    3 & 3 & L & S & 2.051 & 3.112\\
    3 & 3 & L & M & 2.491 & 3.499\\
    3 & 3 & L & L & 2.388 & 3.572\\
\hline
    5 & 3 & S & S & 1.953 & 3.001\\
    5 & 3 & S & M & 2.168 & 3.390\\
    5 & 5 & S & S & 1.954 & 3.000\\
    5 & 5 & S & M & 2.198 & 3.389\\
\end{tabular}
\caption{The effect of different number and size of layers, and optimized regions on the TDDFT-PBE and single point HSE ZPL values. ZPL energies in this table do not include ZPE contribution. While the model describes an additional interstitial $N$ atom,  each small (S) or large (L) layer is defined by 52 and 94 atoms, respectively. In the small (S), medium (M), large (L) relaxation scheme besides the hydrogens the central 4, 13, 37 second-row atoms per layer are taken into account as illustrated in Fig.~\ref{fig:structure_fod}a, respectively. Additionally, ZPL-s with HSE(0.25) functional are  determined on the geometries obtained with PBE. These results are consistently higher around 1 eV regardless of the setup. }
\label{table:size}
\end{table}
The used def2-SVP is a minimal basis set, but using larger basis set would be too expensive. To address this issue, we also study how ZPL changes if larger basis sets are used on the center atoms. In our convergence test, def2-SVPD, def2-TZVP, and def2-TZVPD is applied only on i) the two N-s and two B-s in the center (4 atoms),  on ii) 4 atoms at the center of the 3 middle layer plus the interstitial N (13 atoms) or on iii) 4 atoms of the 5 layer plus the interstitial N (21 atoms). The ZPLs are collected in Table \ref{table:basis}. Using diffuse functions causes less than 0.1 eV change in the ZPL, however the application of TZ quality basis set increases the ZPL from 2.20 eV to 2.28-2.38 eV. It can be also observed that including the outer layers have negligible impact. As a convergent PBE-TDDFT adiabatic ZPL value for the cluster model we obtain 2.38 eV in good agreement with the results of constrained-DFT calculations in a periodic model, see Table~\ref{tab:ZPL}.

\begin{table}
\begin{tabular}{c||c|c|c}
basis & 4 atoms & 13 atoms &  21 atoms  \\
\hline
def2-SVPD &  2.215 & 2.230 & 2.239 \\
def2-TZVP &  2.339 & 2.357 & 2.352 \\
def2-TZVPD & 2.281 & 2.372 & 2.380 \\
 
\end{tabular}
\caption{The effect of larger basis sets on selected atoms on the PBE-TDDFT ZPL. }
\label{table:basis}
\end{table}

\section*{Acknowledgments} 

This research was  supported by the National Research, Development, and Innovation Office of Hungary  within the Quantum Information National Laboratory of Hungary (Grant No. 2022-2.1.1-NL-2022-00004) and within grants FK 135496 and FK 145395.
I.A. acknowledges the Australian Research Council (CE200100010, FT220100053) and the Office of Naval Research Global (N62909-22-1-2028) for the financial support.
V.I. also appreciates support from the Knut and Alice Wallenberg Foundation through WBSQD2 project (Grant No.\ 2018.0071). 
The calculations were performed on resources provided by the Swedish National Infrastructure for Computing (SNIC) at the National Supercomputer Centre (NSC).
We acknowledge KIF\"U for awarding us access to computational resource based in Hungary.

\section*{Data availability}

The data that support the findings of this study are available from the authors upon reasonable request.

\section*{Author contributions}

A.G., R.B, Z.B., and V.I. carried out the first principles calculations. All authors contributed to the manuscript. The work was supervised by V.I. and G.B. 

\section*{Competing interests}

The authors declare no competing interests.


\begin{mcitethebibliography}{57}
\providecommand*\natexlab[1]{#1}
\providecommand*\mciteSetBstSublistMode[1]{}
\providecommand*\mciteSetBstMaxWidthForm[2]{}
\providecommand*\mciteBstWouldAddEndPuncttrue
  {\def\EndOfBibitem{\unskip.}}
\providecommand*\mciteBstWouldAddEndPunctfalse
  {\let\EndOfBibitem\relax}
\providecommand*\mciteSetBstMidEndSepPunct[3]{}
\providecommand*\mciteSetBstSublistLabelBeginEnd[3]{}
\providecommand*\EndOfBibitem{}
\mciteSetBstSublistMode{f}
\mciteSetBstMaxWidthForm{subitem}{(\alph{mcitesubitemcount})}
\mciteSetBstSublistLabelBeginEnd
  {\mcitemaxwidthsubitemform\space}
  {\relax}
  {\relax}

\bibitem[Gisin and Thew(2007)Gisin, and Thew]{Gisin_2007}
Gisin,~N.; Thew,~R. Quantum communication. \emph{Nature Photonics}
  \textbf{2007}, \emph{1}, 165--171\relax
\mciteBstWouldAddEndPuncttrue
\mciteSetBstMidEndSepPunct{\mcitedefaultmidpunct}
{\mcitedefaultendpunct}{\mcitedefaultseppunct}\relax
\EndOfBibitem
\bibitem[Orieux and Diamanti(2016)Orieux, and Diamanti]{Orieux_2016}
Orieux,~A.; Diamanti,~E. Recent advances on integrated quantum communications.
  \emph{Journal of Optics} \textbf{2016}, \emph{18}, 083002\relax
\mciteBstWouldAddEndPuncttrue
\mciteSetBstMidEndSepPunct{\mcitedefaultmidpunct}
{\mcitedefaultendpunct}{\mcitedefaultseppunct}\relax
\EndOfBibitem
\bibitem[Cozzolino \latin{et~al.}(2019)Cozzolino, Da~Lio, Bacco, and
  Oxenløwe]{Cozzolino_2019}
Cozzolino,~D.; Da~Lio,~B.; Bacco,~D.; Oxenløwe,~L.~K. High-Dimensional Quantum
  Communication: Benefits, Progress, and Future Challenges. \emph{Advanced
  Quantum Technologies} \textbf{2019}, \emph{2}, 1900038\relax
\mciteBstWouldAddEndPuncttrue
\mciteSetBstMidEndSepPunct{\mcitedefaultmidpunct}
{\mcitedefaultendpunct}{\mcitedefaultseppunct}\relax
\EndOfBibitem
\bibitem[Kimble(2008)]{kimble_quantum_2008}
Kimble,~H.~J. The quantum internet. \emph{Nature} \textbf{2008}, \emph{453},
  1023--1030\relax
\mciteBstWouldAddEndPuncttrue
\mciteSetBstMidEndSepPunct{\mcitedefaultmidpunct}
{\mcitedefaultendpunct}{\mcitedefaultseppunct}\relax
\EndOfBibitem
\bibitem[Wehner \latin{et~al.}(2018)Wehner, Elkouss, and
  Hanson]{wehner_quantum_2018}
Wehner,~S.; Elkouss,~D.; Hanson,~R. Quantum internet: A vision for the road
  ahead. \emph{Science} \textbf{2018}, \emph{362}\relax
\mciteBstWouldAddEndPuncttrue
\mciteSetBstMidEndSepPunct{\mcitedefaultmidpunct}
{\mcitedefaultendpunct}{\mcitedefaultseppunct}\relax
\EndOfBibitem
\bibitem[Pelucchi \latin{et~al.}(2022)Pelucchi, Fagas, Aharonovich, Englund,
  Figueroa, Gong, Hannes, Liu, Lu, Matsuda, Pan, Schreck, Sciarrino,
  Silberhorn, Wang, and Jöns]{pelucchi_potential_2022}
Pelucchi,~E. \latin{et~al.}  The potential and global outlook of integrated
  photonics for quantum technologies. \emph{Nature Reviews Physics}
  \textbf{2022}, \emph{4}, 194--208, Number: 3 Publisher: Nature Publishing
  Group\relax
\mciteBstWouldAddEndPuncttrue
\mciteSetBstMidEndSepPunct{\mcitedefaultmidpunct}
{\mcitedefaultendpunct}{\mcitedefaultseppunct}\relax
\EndOfBibitem
\bibitem[Caldwell \latin{et~al.}(2019)Caldwell, Aharonovich, Cassabois, Edgar,
  Gil, and Basov]{caldwell_photonics_2019}
Caldwell,~J.~D.; Aharonovich,~I.; Cassabois,~G.; Edgar,~J.~H.; Gil,~B.;
  Basov,~D.~N. Photonics with hexagonal boron nitride. \emph{Nature Reviews
  Materials} \textbf{2019}, \emph{4}, 552--567\relax
\mciteBstWouldAddEndPuncttrue
\mciteSetBstMidEndSepPunct{\mcitedefaultmidpunct}
{\mcitedefaultendpunct}{\mcitedefaultseppunct}\relax
\EndOfBibitem
\bibitem[Aharonovich \latin{et~al.}(2022)Aharonovich, Tetienne, and
  Toth]{aharonovich_quantum_2022}
Aharonovich,~I.; Tetienne,~J.-P.; Toth,~M. Quantum Emitters in Hexagonal Boron
  Nitride. \emph{Nano Letters} \textbf{2022}, \emph{22}, 9227--9235\relax
\mciteBstWouldAddEndPuncttrue
\mciteSetBstMidEndSepPunct{\mcitedefaultmidpunct}
{\mcitedefaultendpunct}{\mcitedefaultseppunct}\relax
\EndOfBibitem
\bibitem[Kubanek(2022)]{kubanek_coherent_2022}
Kubanek,~A. Coherent Quantum Emitters in Hexagonal Boron Nitride.
  \emph{Advanced Quantum Technologies} \textbf{2022}, \emph{5}, 2200009\relax
\mciteBstWouldAddEndPuncttrue
\mciteSetBstMidEndSepPunct{\mcitedefaultmidpunct}
{\mcitedefaultendpunct}{\mcitedefaultseppunct}\relax
\EndOfBibitem
\bibitem[Al-Juboori \latin{et~al.}()Al-Juboori, Zeng, Nguyen, Ai, Laucht,
  Solntsev, Toth, Malaney, and Aharonovich]{al-juboori_quantum_2023}
Al-Juboori,~A.; Zeng,~H. Z.~J.; Nguyen,~M. A.~P.; Ai,~X.; Laucht,~A.;
  Solntsev,~A.; Toth,~M.; Malaney,~R.; Aharonovich,~I. Quantum Key Distribution
  Using an Integrated Quantum Emitter in Hexagonal Boron Nitride.
  \url{http://arxiv.org/abs/2302.06212}\relax
\mciteBstWouldAddEndPuncttrue
\mciteSetBstMidEndSepPunct{\mcitedefaultmidpunct}
{\mcitedefaultendpunct}{\mcitedefaultseppunct}\relax
\EndOfBibitem
\bibitem[Shevitski \latin{et~al.}(2019)Shevitski, Gilbert, Chen, Kastl,
  Barnard, Wong, Ogletree, Watanabe, Taniguchi, Zettl, and
  Aloni]{shevitski_blue-light-emitting_2019}
Shevitski,~B.; Gilbert,~S.~M.; Chen,~C.~T.; Kastl,~C.; Barnard,~E.~S.;
  Wong,~E.; Ogletree,~D.~F.; Watanabe,~K.; Taniguchi,~T.; Zettl,~A.; Aloni,~S.
  Blue-light-emitting color centers in high-quality hexagonal boron nitride.
  \emph{Physical Review B} \textbf{2019}, \emph{100}, 155419, Publisher:
  American Physical Society\relax
\mciteBstWouldAddEndPuncttrue
\mciteSetBstMidEndSepPunct{\mcitedefaultmidpunct}
{\mcitedefaultendpunct}{\mcitedefaultseppunct}\relax
\EndOfBibitem
\bibitem[Fournier \latin{et~al.}(2021)Fournier, Plaud, Roux, Pierret,
  Rosticher, Watanabe, Taniguchi, Buil, Quélin, Barjon, Hermier, and
  Delteil]{fournier_position-controlled_2021}
Fournier,~C.; Plaud,~A.; Roux,~S.; Pierret,~A.; Rosticher,~M.; Watanabe,~K.;
  Taniguchi,~T.; Buil,~S.; Quélin,~X.; Barjon,~J.; Hermier,~J.-P.; Delteil,~A.
  Position-controlled quantum emitters with reproducible emission wavelength in
  hexagonal boron nitride. \emph{Nature Communications} \textbf{2021},
  \emph{12}, 3779, Number: 1 Publisher: Nature Publishing Group\relax
\mciteBstWouldAddEndPuncttrue
\mciteSetBstMidEndSepPunct{\mcitedefaultmidpunct}
{\mcitedefaultendpunct}{\mcitedefaultseppunct}\relax
\EndOfBibitem
\bibitem[Liang \latin{et~al.}()Liang, Chen, Loh, Cheng, Chen, Yang, Zhang,
  Watanabe, Taniguchi, Quek, Bosman, Eda, and Bettiol]{liang_blue_2023}
Liang,~H.; Chen,~Y.; Loh,~L.; Cheng,~N. L.~Q.; Chen,~Y.; Yang,~C.; Zhang,~Z.;
  Watanabe,~K.; Taniguchi,~T.; Quek,~S.~Y.; Bosman,~M.; Eda,~G.; Bettiol,~A.
  Blue Quantum Emitters in hexagonal Boron Nitride.
  \url{https://www.researchsquare.com}\relax
\mciteBstWouldAddEndPuncttrue
\mciteSetBstMidEndSepPunct{\mcitedefaultmidpunct}
{\mcitedefaultendpunct}{\mcitedefaultseppunct}\relax
\EndOfBibitem
\bibitem[Horder \latin{et~al.}(2022)Horder, White, Gale, Li, Watanabe,
  Taniguchi, Kianinia, Aharonovich, and Toth]{horder_coherence_2022}
Horder,~J.; White,~S.; Gale,~A.; Li,~C.; Watanabe,~K.; Taniguchi,~T.;
  Kianinia,~M.; Aharonovich,~I.; Toth,~M. Coherence properties of electron beam
  activated emitters in hexagonal boron nitride under resonant excitation.
  \emph{Physical Review Applied} \textbf{2022}, \emph{18}, 064021\relax
\mciteBstWouldAddEndPuncttrue
\mciteSetBstMidEndSepPunct{\mcitedefaultmidpunct}
{\mcitedefaultendpunct}{\mcitedefaultseppunct}\relax
\EndOfBibitem
\bibitem[Zhigulin \latin{et~al.}()Zhigulin, Horder, Ivady, White, Gale, Li,
  Lobo, Toth, Aharonovich, and Kianinia]{zhigulin_stark_2022}
Zhigulin,~I.; Horder,~J.; Ivady,~V.; White,~S. J.~U.; Gale,~A.; Li,~C.;
  Lobo,~C.~J.; Toth,~M.; Aharonovich,~I.; Kianinia,~M. Stark effect of quantum
  blue emitters in {hBN}. \url{http://arxiv.org/abs/2208.00600}\relax
\mciteBstWouldAddEndPuncttrue
\mciteSetBstMidEndSepPunct{\mcitedefaultmidpunct}
{\mcitedefaultendpunct}{\mcitedefaultseppunct}\relax
\EndOfBibitem
\bibitem[Gale \latin{et~al.}(2022)Gale, Li, Chen, Watanabe, Taniguchi,
  Aharonovich, and Toth]{gale_site-specific_2022}
Gale,~A.; Li,~C.; Chen,~Y.; Watanabe,~K.; Taniguchi,~T.; Aharonovich,~I.;
  Toth,~M. Site-Specific Fabrication of Blue Quantum Emitters in Hexagonal
  Boron Nitride. \emph{{ACS} Photonics} \textbf{2022}, \emph{9}, 2170--2177,
  Publisher: American Chemical Society\relax
\mciteBstWouldAddEndPuncttrue
\mciteSetBstMidEndSepPunct{\mcitedefaultmidpunct}
{\mcitedefaultendpunct}{\mcitedefaultseppunct}\relax
\EndOfBibitem
\bibitem[Mackoit-Sinkevičienė \latin{et~al.}(2019)Mackoit-Sinkevičienė,
  Maciaszek, Van~de Walle, and Alkauskas]{mackoit-sinkeviciene_carbon_2019}
Mackoit-Sinkevičienė,~M.; Maciaszek,~M.; Van~de Walle,~C.~G.; Alkauskas,~A.
  Carbon dimer defect as a source of the 4.1 {eV} luminescence in hexagonal
  boron nitride. \emph{Applied Physics Letters} \textbf{2019}, \emph{115},
  212101, Publisher: American Institute of Physics\relax
\mciteBstWouldAddEndPuncttrue
\mciteSetBstMidEndSepPunct{\mcitedefaultmidpunct}
{\mcitedefaultendpunct}{\mcitedefaultseppunct}\relax
\EndOfBibitem
\bibitem[Hamdi \latin{et~al.}(2020)Hamdi, Thiering, Bodrog, Ivády, and
  Gali]{hamdi_stonewales_2020}
Hamdi,~H.; Thiering,~G.; Bodrog,~Z.; Ivády,~V.; Gali,~A. Stone–Wales defects
  in hexagonal boron nitride as ultraviolet emitters. \emph{npj Computational
  Materials} \textbf{2020}, \emph{6}, 1--7\relax
\mciteBstWouldAddEndPuncttrue
\mciteSetBstMidEndSepPunct{\mcitedefaultmidpunct}
{\mcitedefaultendpunct}{\mcitedefaultseppunct}\relax
\EndOfBibitem
\bibitem[Li \latin{et~al.}(2022)Li, Pershin, Thiering, Udvarhelyi, and
  Gali]{li_ultraviolet_2022}
Li,~S.; Pershin,~A.; Thiering,~G.; Udvarhelyi,~P.; Gali,~A. Ultraviolet Quantum
  Emitters in Hexagonal Boron Nitride from Carbon Clusters. \emph{The Journal
  of Physical Chemistry Letters} \textbf{2022}, \emph{13}, 3150--3157\relax
\mciteBstWouldAddEndPuncttrue
\mciteSetBstMidEndSepPunct{\mcitedefaultmidpunct}
{\mcitedefaultendpunct}{\mcitedefaultseppunct}\relax
\EndOfBibitem
\bibitem[Zhigulin \latin{et~al.}()Zhigulin, Yamamura, Ivády, Gale, Horder,
  Lobo, Kianinia, Toth, and Aharonovich]{zhigulin_photophysics_2023}
Zhigulin,~I.; Yamamura,~K.; Ivády,~V.; Gale,~A.; Horder,~J.; Lobo,~C.~J.;
  Kianinia,~M.; Toth,~M.; Aharonovich,~I. Photophysics of blue quantum emitters
  in hexagonal Boron Nitride. \url{http://arxiv.org/abs/2301.04269}\relax
\mciteBstWouldAddEndPuncttrue
\mciteSetBstMidEndSepPunct{\mcitedefaultmidpunct}
{\mcitedefaultendpunct}{\mcitedefaultseppunct}\relax
\EndOfBibitem
\bibitem[Wang \latin{et~al.}(2014)Wang, Liu, He, Yang, and Ma]{Wang2014def}
Wang,~V.; Liu,~R.-J.; He,~H.-P.; Yang,~C.-M.; Ma,~L. Hybrid functional with
  semi-empirical van der Waals study of native defects in hexagonal {BN}.
  \emph{Solid State Communications} \textbf{2014}, \emph{177}, 74--79\relax
\mciteBstWouldAddEndPuncttrue
\mciteSetBstMidEndSepPunct{\mcitedefaultmidpunct}
{\mcitedefaultendpunct}{\mcitedefaultseppunct}\relax
\EndOfBibitem
\bibitem[Weston \latin{et~al.}(2018)Weston, Wickramaratne, Mackoit, Alkauskas,
  and Van~de Walle]{weston_native_2018}
Weston,~L.; Wickramaratne,~D.; Mackoit,~M.; Alkauskas,~A.; Van~de Walle,~C.~G.
  Native point defects and impurities in hexagonal boron nitride.
  \emph{Physical Review B} \textbf{2018}, \emph{97}, 214104\relax
\mciteBstWouldAddEndPuncttrue
\mciteSetBstMidEndSepPunct{\mcitedefaultmidpunct}
{\mcitedefaultendpunct}{\mcitedefaultseppunct}\relax
\EndOfBibitem
\bibitem[Strand \latin{et~al.}(2019)Strand, Larcher, and Shluger]{Strand2019}
Strand,~J.; Larcher,~L.; Shluger,~A.~L. Properties of intrinsic point defects
  and dimers in hexagonal boron nitride. \emph{Journal of Physics: Condensed
  Matter} \textbf{2019}, \emph{32}, 055706\relax
\mciteBstWouldAddEndPuncttrue
\mciteSetBstMidEndSepPunct{\mcitedefaultmidpunct}
{\mcitedefaultendpunct}{\mcitedefaultseppunct}\relax
\EndOfBibitem
\bibitem[Khorasani \latin{et~al.}(2021)Khorasani, Frauenheim, Aradi, and
  Deák]{khorasani_identification_2021}
Khorasani,~E.; Frauenheim,~T.; Aradi,~B.; Deák,~P. Identification of the
  Nitrogen Interstitial as Origin of the 3.1 {eV} Photoluminescence Band in
  Hexagonal Boron Nitride. \emph{physica status solidi (b)} \textbf{2021},
  \emph{258}, 2100031\relax
\mciteBstWouldAddEndPuncttrue
\mciteSetBstMidEndSepPunct{\mcitedefaultmidpunct}
{\mcitedefaultendpunct}{\mcitedefaultseppunct}\relax
\EndOfBibitem
\bibitem[Bhang \latin{et~al.}(2021)Bhang, Ma, Yim, Galli, and
  Seo]{bhang_first-principles_2021}
Bhang,~J.; Ma,~H.; Yim,~D.; Galli,~G.; Seo,~H. First-Principles Predictions of
  Out-of-Plane Group {IV} and V Dimers as High-Symmetry, High-Spin Defects in
  Hexagonal Boron Nitride. \emph{{ACS} Applied Materials \& Interfaces}
  \textbf{2021}, \emph{13}, 45768--45777, Publisher: American Chemical
  Society\relax
\mciteBstWouldAddEndPuncttrue
\mciteSetBstMidEndSepPunct{\mcitedefaultmidpunct}
{\mcitedefaultendpunct}{\mcitedefaultseppunct}\relax
\EndOfBibitem
\bibitem[Lany and Zunger(2009)Lany, and Zunger]{lany_polaronic_2009}
Lany,~S.; Zunger,~A. Polaronic hole localization and multiple hole binding of
  acceptors in oxide wide-gap semiconductors. \emph{Phys. Rev. B}
  \textbf{2009}, \emph{80}, 085202\relax
\mciteBstWouldAddEndPuncttrue
\mciteSetBstMidEndSepPunct{\mcitedefaultmidpunct}
{\mcitedefaultendpunct}{\mcitedefaultseppunct}\relax
\EndOfBibitem
\bibitem[Lany and Zunger(2010)Lany, and Zunger]{lany_generalized_2010}
Lany,~S.; Zunger,~A. Generalized Koopmans density functional calculations
  reveal the deep acceptor state of ${\text{N}}_{\text{O}}$ in ZnO. \emph{Phys.
  Rev. B} \textbf{2010}, \emph{81}, 205209\relax
\mciteBstWouldAddEndPuncttrue
\mciteSetBstMidEndSepPunct{\mcitedefaultmidpunct}
{\mcitedefaultendpunct}{\mcitedefaultseppunct}\relax
\EndOfBibitem
\bibitem[Iv\'ady \latin{et~al.}(2013)Iv\'ady, Abrikosov, Janz\'en, and
  Gali]{ivady_role_2013}
Iv\'ady,~V.; Abrikosov,~I.~A.; Janz\'en,~E.; Gali,~A. Role of screening in the
  density functional applied to transition-metal defects in semiconductors.
  \emph{Phys. Rev. B} \textbf{2013}, \emph{87}, 205201\relax
\mciteBstWouldAddEndPuncttrue
\mciteSetBstMidEndSepPunct{\mcitedefaultmidpunct}
{\mcitedefaultendpunct}{\mcitedefaultseppunct}\relax
\EndOfBibitem
\bibitem[Grimme and Hansen(2015)Grimme, and Hansen]{grimme2015practicable}
Grimme,~S.; Hansen,~A. A practicable real-space measure and visualization of
  static electron-correlation effects. \emph{Angewandte Chemie International
  Edition} \textbf{2015}, \emph{54}, 12308--12313\relax
\mciteBstWouldAddEndPuncttrue
\mciteSetBstMidEndSepPunct{\mcitedefaultmidpunct}
{\mcitedefaultendpunct}{\mcitedefaultseppunct}\relax
\EndOfBibitem
\bibitem[Bauer \latin{et~al.}(2017)Bauer, Hansen, and
  Grimme]{bauer2017fractional}
Bauer,~C.~A.; Hansen,~A.; Grimme,~S. The fractional occupation number weighted
  density as a versatile analysis tool for molecules with a complicated
  electronic structure. \emph{Chemistry--A European Journal} \textbf{2017},
  \emph{23}, 6150--6164\relax
\mciteBstWouldAddEndPuncttrue
\mciteSetBstMidEndSepPunct{\mcitedefaultmidpunct}
{\mcitedefaultendpunct}{\mcitedefaultseppunct}\relax
\EndOfBibitem
\bibitem[Gouveia and Coutinho(2019)Gouveia, and Coutinho]{Gouveia2019}
Gouveia,~J.~D.; Coutinho,~J. Can we rely on hybrid-{DFT} energies of
  solid-state problems with local-{DFT} geometries? \emph{Electronic Structure}
  \textbf{2019}, \emph{1}, 015008\relax
\mciteBstWouldAddEndPuncttrue
\mciteSetBstMidEndSepPunct{\mcitedefaultmidpunct}
{\mcitedefaultendpunct}{\mcitedefaultseppunct}\relax
\EndOfBibitem
\bibitem[Perdew \latin{et~al.}(1996)Perdew, Burke, and Ernzerhof]{PBE}
Perdew,~J.~P.; Burke,~K.; Ernzerhof,~M. Generalized Gradient Approximation Made
  Simple. \emph{Phys. Rev. Lett.} \textbf{1996}, \emph{77}, 3865--3868\relax
\mciteBstWouldAddEndPuncttrue
\mciteSetBstMidEndSepPunct{\mcitedefaultmidpunct}
{\mcitedefaultendpunct}{\mcitedefaultseppunct}\relax
\EndOfBibitem
\bibitem[Gilbert \latin{et~al.}(2019)Gilbert, Pham, Dogan, Oh, Shevitski,
  Schumm, Liu, Ercius, Aloni, Cohen, and Zettl]{Gilbert2019}
Gilbert,~S.~M.; Pham,~T.; Dogan,~M.; Oh,~S.; Shevitski,~B.; Schumm,~G.;
  Liu,~S.; Ercius,~P.; Aloni,~S.; Cohen,~M.~L.; Zettl,~A. Alternative stacking
  sequences in hexagonal boron nitride. \emph{2D Materials} \textbf{2019},
  \emph{6}, 021006\relax
\mciteBstWouldAddEndPuncttrue
\mciteSetBstMidEndSepPunct{\mcitedefaultmidpunct}
{\mcitedefaultendpunct}{\mcitedefaultseppunct}\relax
\EndOfBibitem
\bibitem[Ivady \latin{et~al.}(2020)Ivady, Barcza, Thiering, Li, Hamdi, Chou,
  Legeza, and Gali]{ivady_ab_2020}
Ivady,~V.; Barcza,~G.; Thiering,~G.; Li,~S.; Hamdi,~H.; Chou,~J.-P.;
  Legeza,~O.; Gali,~A. Ab initio theory of the negatively charged boron vacancy
  qubit in hexagonal boron nitride. \emph{npj Computational Materials}
  \textbf{2020}, \emph{6}, 1--6, Number: 1 Publisher: Nature Publishing
  Group\relax
\mciteBstWouldAddEndPuncttrue
\mciteSetBstMidEndSepPunct{\mcitedefaultmidpunct}
{\mcitedefaultendpunct}{\mcitedefaultseppunct}\relax
\EndOfBibitem
\bibitem[Babar \latin{et~al.}()Babar, Barcza, Pershin, Park, Lindvall,
  Thiering, Legeza, Warner, Abrikosov, Gali, and Ivády]{babar_quantum_2021}
Babar,~R.; Barcza,~G.; Pershin,~A.; Park,~H.; Lindvall,~O.~B.; Thiering,~G.;
  Legeza,~O.; Warner,~J.~H.; Abrikosov,~I.~A.; Gali,~A.; Ivády,~V. Quantum
  sensor in a single layer van der Waals material.
  \url{http://arxiv.org/abs/2111.09589}\relax
\mciteBstWouldAddEndPuncttrue
\mciteSetBstMidEndSepPunct{\mcitedefaultmidpunct}
{\mcitedefaultendpunct}{\mcitedefaultseppunct}\relax
\EndOfBibitem
\bibitem[Benedek \latin{et~al.}()Benedek, Babar, Ganyecz, Szilvasi, Legeza,
  Barcza, and Ivady]{benedek_symmetric_2023}
Benedek,~Z.; Babar,~R.; Ganyecz,~A.; Szilvasi,~T.; Legeza,~O.; Barcza,~G.;
  Ivady,~V. Symmetric carbon tetramers forming chemically stable spin qubits in
  {hBN}. \url{http://arxiv.org/abs/2303.14110}\relax
\mciteBstWouldAddEndPuncttrue
\mciteSetBstMidEndSepPunct{\mcitedefaultmidpunct}
{\mcitedefaultendpunct}{\mcitedefaultseppunct}\relax
\EndOfBibitem
\bibitem[Angeli \latin{et~al.}(2002)Angeli, Cimiraglia, and
  Malrieu]{Angeli-2001a}
Angeli,~C.; Cimiraglia,~R.; Malrieu,~J.-P. n-electron valence state
  perturbation theory: A spinless formulation and an efficient implementation
  of the strongly contracted and of the partially contracted variants.
  \emph{The Journal of Chemical Physics} \textbf{2002}, \emph{117},
  9138--9153\relax
\mciteBstWouldAddEndPuncttrue
\mciteSetBstMidEndSepPunct{\mcitedefaultmidpunct}
{\mcitedefaultendpunct}{\mcitedefaultseppunct}\relax
\EndOfBibitem
\bibitem[Heyd \latin{et~al.}(2006)Heyd, Scuseria, and Ernzerhof]{HSE06}
Heyd,~J.; Scuseria,~G.~E.; Ernzerhof,~M. Erratum: “Hybrid functionals based
  on a screened Coulomb potential” [ J. Chem. Phys. 118, 8207 (2003) ].
  \emph{J. Chem. Phys.} \textbf{2006}, \emph{124}, 219906\relax
\mciteBstWouldAddEndPuncttrue
\mciteSetBstMidEndSepPunct{\mcitedefaultmidpunct}
{\mcitedefaultendpunct}{\mcitedefaultseppunct}\relax
\EndOfBibitem
\bibitem[Perdew \latin{et~al.}(1982)Perdew, Parr, Levy, and Balduz]{Perdew91}
Perdew,~J.~P.; Parr,~R.~G.; Levy,~M.; Balduz,~J.~L. Density-Functional Theory
  for Fractional Particle Number: Derivative Discontinuities of the Energy.
  \emph{Phys. Rev. Lett.} \textbf{1982}, \emph{49}, 1691--1694\relax
\mciteBstWouldAddEndPuncttrue
\mciteSetBstMidEndSepPunct{\mcitedefaultmidpunct}
{\mcitedefaultendpunct}{\mcitedefaultseppunct}\relax
\EndOfBibitem
\bibitem[Perdew and Levy(1997)Perdew, and Levy]{Perdew97}
Perdew,~J.~P.; Levy,~M. Comment on “Significance of the highest occupied
  Kohn-Sham eigenvalue”. \emph{Phys. Rev. B} \textbf{1997}, \emph{56},
  16021--16028\relax
\mciteBstWouldAddEndPuncttrue
\mciteSetBstMidEndSepPunct{\mcitedefaultmidpunct}
{\mcitedefaultendpunct}{\mcitedefaultseppunct}\relax
\EndOfBibitem
\bibitem[Almbladh and von Barth(1985)Almbladh, and von Barth]{Almbladh85}
Almbladh,~C.-O.; von Barth,~U. Exact results for the charge and spin densities,
  exchange-correlation potentials, and density-functional eigenvalues.
  \emph{Phys. Rev. B} \textbf{1985}, \emph{31}, 3231--3244\relax
\mciteBstWouldAddEndPuncttrue
\mciteSetBstMidEndSepPunct{\mcitedefaultmidpunct}
{\mcitedefaultendpunct}{\mcitedefaultseppunct}\relax
\EndOfBibitem
\bibitem[Freysoldt \latin{et~al.}(2009)Freysoldt, Neugebauer, and Van~de
  Walle]{Freysoldt}
Freysoldt,~C.; Neugebauer,~J.; Van~de Walle,~C.~G. Fully \textit{Ab Initio}
  Finite-Size Corrections for Charged-Defect Supercell Calculations.
  \emph{Phys. Rev. Lett.} \textbf{2009}, \emph{102}, 016402\relax
\mciteBstWouldAddEndPuncttrue
\mciteSetBstMidEndSepPunct{\mcitedefaultmidpunct}
{\mcitedefaultendpunct}{\mcitedefaultseppunct}\relax
\EndOfBibitem
\bibitem[Lany and Zunger(2008)Lany, and Zunger]{lany_assessment_2008}
Lany,~S.; Zunger,~A. Assessment of correction methods for the band-gap problem
  and for finite-size effects in supercell defect calculations: Case studies
  for ZnO and GaAs. \emph{Phys. Rev. B} \textbf{2008}, \emph{78}, 235104\relax
\mciteBstWouldAddEndPuncttrue
\mciteSetBstMidEndSepPunct{\mcitedefaultmidpunct}
{\mcitedefaultendpunct}{\mcitedefaultseppunct}\relax
\EndOfBibitem
\bibitem[Wang \latin{et~al.}(2015)Wang, Han, Li, Xie, Chen, Tian, West, Sun,
  and Zhang]{wang_determination_2015}
Wang,~D.; Han,~D.; Li,~X.-B.; Xie,~S.-Y.; Chen,~N.-K.; Tian,~W.~Q.; West,~D.;
  Sun,~H.-B.; Zhang,~S. Determination of Formation and Ionization Energies of
  Charged Defects in Two-Dimensional Materials. \emph{Physical Review Letters}
  \textbf{2015}, \emph{114}, 196801\relax
\mciteBstWouldAddEndPuncttrue
\mciteSetBstMidEndSepPunct{\mcitedefaultmidpunct}
{\mcitedefaultendpunct}{\mcitedefaultseppunct}\relax
\EndOfBibitem
\bibitem[Komsa \latin{et~al.}(2014)Komsa, Berseneva, Krasheninnikov, and
  Nieminen]{komsa_charged_2014}
Komsa,~H.-P.; Berseneva,~N.; Krasheninnikov,~A.~V.; Nieminen,~R.~M. Charged
  Point Defects in the Flatland: Accurate Formation Energy Calculations in
  Two-Dimensional Materials. \emph{Phys. Rev. X} \textbf{2014}, \emph{4},
  031044\relax
\mciteBstWouldAddEndPuncttrue
\mciteSetBstMidEndSepPunct{\mcitedefaultmidpunct}
{\mcitedefaultendpunct}{\mcitedefaultseppunct}\relax
\EndOfBibitem
\bibitem[Weigend and Ahlrichs(2005)Weigend, and Ahlrichs]{weigend2005balanced}
Weigend,~F.; Ahlrichs,~R. Balanced basis sets of split valence, triple zeta
  valence and quadruple zeta valence quality for H to Rn: Design and assessment
  of accuracy. \emph{Physical Chemistry Chemical Physics} \textbf{2005},
  \emph{7}, 3297--3305\relax
\mciteBstWouldAddEndPuncttrue
\mciteSetBstMidEndSepPunct{\mcitedefaultmidpunct}
{\mcitedefaultendpunct}{\mcitedefaultseppunct}\relax
\EndOfBibitem
\bibitem[Perdew \latin{et~al.}(1996)Perdew, Burke, and
  Ernzerhof]{perdew1996generalized}
Perdew,~J.~P.; Burke,~K.; Ernzerhof,~M. Generalized gradient approximation made
  simple. \emph{Physical review letters} \textbf{1996}, \emph{77}, 3865\relax
\mciteBstWouldAddEndPuncttrue
\mciteSetBstMidEndSepPunct{\mcitedefaultmidpunct}
{\mcitedefaultendpunct}{\mcitedefaultseppunct}\relax
\EndOfBibitem
\bibitem[Petrenko and Neese(2012)Petrenko, and Neese]{petrenko2012efficient}
Petrenko,~T.; Neese,~F. Efficient and automatic calculation of optical band
  shapes and resonance Raman spectra for larger molecules within the
  independent mode displaced harmonic oscillator model. \emph{The Journal of
  Chemical Physics} \textbf{2012}, \emph{137}, 234107\relax
\mciteBstWouldAddEndPuncttrue
\mciteSetBstMidEndSepPunct{\mcitedefaultmidpunct}
{\mcitedefaultendpunct}{\mcitedefaultseppunct}\relax
\EndOfBibitem
\bibitem[Ferrer and Santoro(2012)Ferrer, and Santoro]{ferrer2012comparison}
Ferrer,~F. J.~A.; Santoro,~F. Comparison of vertical and adiabatic harmonic
  approaches for the calculation of the vibrational structure of electronic
  spectra. \emph{Physical Chemistry Chemical Physics} \textbf{2012}, \emph{14},
  13549--13563\relax
\mciteBstWouldAddEndPuncttrue
\mciteSetBstMidEndSepPunct{\mcitedefaultmidpunct}
{\mcitedefaultendpunct}{\mcitedefaultseppunct}\relax
\EndOfBibitem
\bibitem[Noh \latin{et~al.}(2018)Noh, Choi, Kim, Im, Kim, Seo, and
  Lee]{noh_stark_2018}
Noh,~G.; Choi,~D.; Kim,~J.-H.; Im,~D.-G.; Kim,~Y.-H.; Seo,~H.; Lee,~J. Stark
  Tuning of Single-Photon Emitters in Hexagonal Boron Nitride. \emph{Nano
  Letters} \textbf{2018}, \emph{18}, 4710--4715, Publisher: American Chemical
  Society\relax
\mciteBstWouldAddEndPuncttrue
\mciteSetBstMidEndSepPunct{\mcitedefaultmidpunct}
{\mcitedefaultendpunct}{\mcitedefaultseppunct}\relax
\EndOfBibitem
\bibitem[Nikolay \latin{et~al.}(2019)Nikolay, Mendelson, Sadzak, Böhm, Tran,
  Sontheimer, Aharonovich, and Benson]{nikolay_very_2019}
Nikolay,~N.; Mendelson,~N.; Sadzak,~N.; Böhm,~F.; Tran,~T.~T.; Sontheimer,~B.;
  Aharonovich,~I.; Benson,~O. Very Large and Reversible Stark-Shift Tuning of
  Single Emitters in Layered Hexagonal Boron Nitride. \emph{Physical Review
  Applied} \textbf{2019}, \emph{11}, 041001, Publisher: American Physical
  Society\relax
\mciteBstWouldAddEndPuncttrue
\mciteSetBstMidEndSepPunct{\mcitedefaultmidpunct}
{\mcitedefaultendpunct}{\mcitedefaultseppunct}\relax
\EndOfBibitem
\bibitem[Kresse and Furthm\"uller(1996)Kresse, and Furthm\"uller]{VASP2}
Kresse,~G.; Furthm\"uller,~J. Efficient iterative schemes for \textit{ab
  initio} total-energy calculations using a plane-wave basis set. \emph{Phys.
  Rev. B} \textbf{1996}, \emph{54}, 11169--11186\relax
\mciteBstWouldAddEndPuncttrue
\mciteSetBstMidEndSepPunct{\mcitedefaultmidpunct}
{\mcitedefaultendpunct}{\mcitedefaultseppunct}\relax
\EndOfBibitem
\bibitem[Bl\"ochl(1994)]{PAW}
Bl\"ochl,~P.~E. Projector augmented-wave method. \emph{Phys. Rev. B}
  \textbf{1994}, \emph{50}, 17953--17979\relax
\mciteBstWouldAddEndPuncttrue
\mciteSetBstMidEndSepPunct{\mcitedefaultmidpunct}
{\mcitedefaultendpunct}{\mcitedefaultseppunct}\relax
\EndOfBibitem
\bibitem[Grimme \latin{et~al.}(2010)Grimme, Antony, Ehrlich, and
  Krieg]{Grimme2010}
Grimme,~S.; Antony,~J.; Ehrlich,~S.; Krieg,~H. A consistent and accurate ab
  initio parametrization of density functional dispersion correction (DFT-D)
  for the 94 elements H-Pu. \emph{The Journal of Chemical Physics}
  \textbf{2010}, \emph{132}, 154104\relax
\mciteBstWouldAddEndPuncttrue
\mciteSetBstMidEndSepPunct{\mcitedefaultmidpunct}
{\mcitedefaultendpunct}{\mcitedefaultseppunct}\relax
\EndOfBibitem
\bibitem[Grimme \latin{et~al.}(2011)Grimme, Ehrlich, and Goerigk]{Grimme2011}
Grimme,~S.; Ehrlich,~S.; Goerigk,~L. Effect of the damping function in
  dispersion corrected density functional theory. \emph{Journal of
  Computational Chemistry} \textbf{2011}, \emph{32}, 1456--1465\relax
\mciteBstWouldAddEndPuncttrue
\mciteSetBstMidEndSepPunct{\mcitedefaultmidpunct}
{\mcitedefaultendpunct}{\mcitedefaultseppunct}\relax
\EndOfBibitem
\bibitem[Paszkowicz \latin{et~al.}(2002)Paszkowicz, Pelka, Knapp, Szyszko, and
  Podsiadlo]{BNlattice}
Paszkowicz,~W.; Pelka,~J.; Knapp,~M.; Szyszko,~T.; Podsiadlo,~S. Lattice
  parameters and anisotropic thermal expansion of hexagonal boron nitride in
  the 10-297.5K temperature range. \emph{Applied Physics A} \textbf{2002},
  \emph{75}, 431--435\relax
\mciteBstWouldAddEndPuncttrue
\mciteSetBstMidEndSepPunct{\mcitedefaultmidpunct}
{\mcitedefaultendpunct}{\mcitedefaultseppunct}\relax
\EndOfBibitem
\bibitem[Neese(2022)]{neese2022software}
Neese,~F. Software update: The ORCA program system—Version 5.0. \emph{Wiley
  Interdisciplinary Reviews: Computational Molecular Science} \textbf{2022},
  \emph{12}, e1606\relax
\mciteBstWouldAddEndPuncttrue
\mciteSetBstMidEndSepPunct{\mcitedefaultmidpunct}
{\mcitedefaultendpunct}{\mcitedefaultseppunct}\relax
\EndOfBibitem
\end{mcitethebibliography}
\providecommand{\latin}[1]{#1}
\makeatletter
\providecommand{\doi}
  {\begingroup\let\do\@makeother\dospecials
  \catcode`\{=1 \catcode`\}=2 \doi@aux}
\providecommand{\doi@aux}[1]{\endgroup\texttt{#1}}
\makeatother
\providecommand*\mcitethebibliography{\thebibliography}
\csname @ifundefined\endcsname{endmcitethebibliography}
  {\let\endmcitethebibliography\endthebibliography}{}

\end{document}